\documentclass[a4paper,12pt]{amsart}

\usepackage{amssymb,mathtools,lmodern}

\usepackage[colorlinks=true,linkcolor=blue,citecolor=blue]{hyperref}

\newcommand{\pprime}{\prime\prime}

\DeclareMathOperator{\Ai}{Ai}
\DeclareMathOperator{\Bi}{Bi}

\newcommand{\he}{\mathfrak{H}}
\newcommand{\se}{\mathfrak{S}}
\newcommand{\dr}{\dot{\rho}_{0}}
\newcommand{\dt}{\dot{\tau}_{0}}
\newcommand{\rg}{\rho_{g}}
\newcommand{\K}{K_{0}}
\newcommand{\w}{w_{0}}

\mathtoolsset{showonlyrefs}

\begin{document}
\title[Variable eddy viscosity from ageostrophic flows]{Variable eddy viscosities in the atmospheric boundary layer from ageostrophic wind-speed profiles}

\author{Tony Lyons}

\address{Department of Computing \& Mathematics, Waterford Institute of Technology,\newline Waterford, Ireland} 
\email{tlyons@wit.ie}


\maketitle

\begin{abstract}
We generate explicit height-dependent eddy viscosity coefficients in the Ekman layer from convex wind speed profiles. The solutions we obtain are parameterized in terms of the relative deflection angle between the wind directions at the top and bottom of the flow, as well as the geostrophic wind speed and a velocity scale we interpret as the transfer rate of horizontal momentum in the vertical direction. The solutions may be used to infer the thickness of the Ekman layer for a variety of deflection angles different from deflection angle of the classic Ekman spiral.
\end{abstract}

\section{Introduction}
A recent reformulation of the governing equations describing transport of horizontal momentum in the planetary boundary layer (PBL)  is used in this paper to obtain eddy viscosity coefficients from Ekman flows in this atmospheric layer. The reformulation of these governing equations was first proposed in \cite{CJ2019} as part of a broader investigation of Ekman flows in the PBL. Surprisingly, the reformulation of the linear system of  governing equations as a system of coupled, nonlinear ordinary differential equations allows one  to obtain the eddy viscosity profile associated with the wind speed profile in the atmospheric boundary layer. This is in contrast with many other approaches which are often used to deduce the ageostrophic wind velocity starting with a specific eddy viscosity profile. The aim of the work here is to consider a general class of wind speed profiles which decay exponentially with height, from which we deduce separable ordinary differential equations to describe the wind speed and direction.  In turn, the observations made in \cite{CJ2019} will allow us to derive the vertical profile of the eddy viscosity coefficient. In the following, we show that this procedure allows us to recover several wind speed profiles investigated in \cite{CJ2019}, and to analyse several other wind velocity profiles and their associated eddy viscosity coefficients.

Within the Ekman layer turbulent airflow is understood as a balance between pressure gradients, the Coriolis force and eddy viscosity \cite{Hol2004,Ped2013}. The eddy viscosity is governed by the flow structure of the Ekman layer and is not a property of the fluid itself. Classical Ekman theory assumes an eddy viscosity coefficient of the form $K(z)=K_{0}$, where $K_{0}$ is some constant. Under these assumptions the large scale atmospheric flow in this atmospheric layer is described by the Ekman spiral, cf. \cite{Ekm1905}. This solution was first derived as a model of wind-driven surface ocean currents (see \cite{CJ2019a} for a generalisation to shallow-water Ekman flows described using spherical coordinates valid at mid latitudes and near the equator). The wind in the Ekman layer may be decomposed into contributions from geostrophic and ageostrophic wind components with a relative deflection of $45^\circ$ between the wind direction at the top and bottom. Ascending the Ekman layer, the wind direction always rotates clockwise in the northern hemisphere and its speed increases monotonically, until it aligns with the geostrophic wind in the free atmosphere above the PBL.

At mid-latitudes the absence of observational data in support of the classical Ekman flow indicates the unsuitability of the constant $K$ model, instead indicating eddy viscosities which vary with vertical height. There are numerous models for the vertical profile of $K(z)$, with many featuring rapid vertical gradients near the base of the PBL \cite{Hol2004}, while others incorporate a steady linear increase in the lower third of the Ekman layer followed by an exponential decline towards the free atmosphere. The widely used model due to O'Brien cf. \cite{OBr1970} as well as a similarly shaped profile due to Acker et al \cite{ABCD1973} both feature profiles with slow growth near the base of the Ekman layer with a turning point followed by rapid decay towards the geostrophic layer above. In contrast the works \cite{Mil1994,Mads1977} investigate flows obtained from viscosity profiles which decay steadily all along the Ekman layer. The models considered in \cite{Gri1995,BG1998,Tan2001,ZT2002} analyse  Ekman flows using perturbative WKB approximations incorporating monotonic eddy viscosity coefficients which vary slowly with height in the PBL. In other cases, the eddy diffusion coefficient is deduced numerically, for instance Deardorff \cite{Dea1970} has numerically integrated the nonlinear equations of motion for mesoscale flows in the planetary boundary layer and derived the eddy viscosity distribution \emph{a posteriori} from these flows.

A notable result from \cite{CJ2019} is that instead of presupposing the vertical dependence of the eddy viscosity, one may start from a given wind speed profile and then derive the associated eddy viscosity function. The nonlinear system describing mesoscale flows proposed in \cite{CJ2019} is based on a re-parameterisation of the height variable combined with a reformulation of the governing equations in polar form, which is then used to obtain the deflection angle and eddy viscosity coefficients associated with a given wind speed profile. In this paper we do not start from explicitly prescribed wind speed profiles, instead we consider a general class of exponentially decaying wind speeds, whose convexity is used to deduce a general class of separable ordinary differential equations governing the height dependence of the wind speed.  It is found that these viscosity profiles are characterised by the value of the eddy viscosity coefficient near the surface boundary layer and the relative deflection between the wind velocity at the top and bottom the Ekman layer. In some cases analytic expressions for the eddy viscosity may be found in terms of the vertical height. In other cases, the monotonic re-parameterisation of the height must be inverted numerically to obtain the vertical dependence of the  eddy viscosity coefficient. In all cases considered it is found that the relative deflection between the wind directions at the top and bottom of the Ekman layer may depart from $45^\circ$ as observed in the classical Ekman spiral. However, the clockwise rotation of the wind direction with increasing height is reproduced in each case. The last section of this paper considers a Riccati type equation governing the exponentially decaying wind speed, and outlines a similar process for generating the vertical profile of the associated eddy viscosity coefficient and wind velocity from this class of equations.

\section{General features of Ekman flows with variable eddy viscosity}

Depending on atmospheric conditions, the Ekman layer may begin between 20 and 100 meters above the surface layer with an upper boundary in excess of 1000 meters and comprises approximately 90\% of earth's atmosphere. The fluid motion in this layer is primarily governed by pressure gradients within the fluid, along with frictional and Coriolis forces. The conventional governing equations for the atmospheric flow generated by this system of forces is given by the system
\begin{equation}\label{eq:Ekman2Component}
\begin{aligned}
 f(u-u_{g})&=\frac{d}{dz}\left(K\frac{dv}{dz}\right)\\
 f(v-v_{g})&=-\frac{d}{dz}\left(K\frac{du}{dz}\right)
\end{aligned}
\end{equation}
with $u$ \& $v$ being the mean wind velocities in the zonal and meridional directions respectively and $K$ is the eddy viscosity in the Ekman layer. The constant $f=2\Omega\sin(\phi)$ is the Coriolis parameter at latitude $\phi$ while $\Omega=7.29\times10^{-5}\mathrm{rad\,s}^{-1}$ is the rotation speed of the earth. The interface between the Prandtl layer and the Ekman layer is denoted by $z=0$, and the no-slip the boundary condition between these two layers is given by
\begin{equation}\label{eq:PrandtlBC}
    (u,v)=(0,0) \text{ at }z=0,
\end{equation}
while the boundary condition
\begin{equation}\label{eq:GeostrophicBC}
    (u,v)\to(u_{g},v_{g})\text{ as }z\to\infty,
\end{equation}
ensures the wind achieves geostrophic balance above the PBL. The solution of the  system \eqref{eq:Ekman2Component}--\eqref{eq:GeostrophicBC} with constant eddy viscosity $K$ is given by
\begin{equation}\label{eq:uv-Solution}
\begin{aligned}
u(z)&=u_{g}-e^{-\gamma z}\left[u_{g}\cos(\gamma z)+v_{g}\sin(\gamma z)\right]\\
v(z)&=v_{g}+e^{-\gamma z}\left[v_{g}\cos(\gamma z)-u_{g}\sin(\gamma z)\right],
\end{aligned}
\end{equation}
where we introduce the parameter
\[\gamma=\sqrt{\frac{f}{2K}}.\]
This is the classic Ekman spiral wherein the wind direction rotates clockwise and wind speed increases monotonically toward the geostrophic wind speed, with increasing height. While the classical Ekman spiral is rarely observed at mid-latitudes, \cite{Rys_etal2016} provides evidence for Ekman spirals from wind-speed field data collected at Dome C on the Antarctic Plateau, where atmospheric conditions may allow for wind patterns resembling the classical Ekman spiral. At mid-latitudes  a non-constant eddy viscosity is more appropriate, making the system \eqref{eq:Ekman2Component} considerably more difficult to analyse, however recent progress has been made in obtaining flows associated with various forms of $K(z)$, cf. \cite{Con2020,DPC2020,Ion2021}.

\subsection{Re-parameterisation of the system}
The solution \eqref{eq:uv-Solution}, when written in complex form is given by
\[(u(z)-u_{g})+i(v(z)-v_{g})=-e^{-(1+i)\gamma z}(u_{g}+iv_{g}).\]
This form of the classical Ekman solution shows that the wind velocity $(u(z),v(z))$ spirals clockwise with increasing height $z$, until it aligns with the geostrophic wind velocity $(u_{g},v_{g})$ in the free atmospheric layer above. The work in \cite{CJ2019} extends the constant $K$ model to more general systems \eqref{eq:Ekman2Component}--\eqref{eq:GeostrophicBC} with varying eddy viscosity profiles $K(z)$,  subject only to the conditions $K:\left[0,\infty\right) \to \left[k_{-},k_{+}\right]$ and $\lim_{z\to\infty}K(z)= k^{*}\in\left[k_{-},k_{+}\right]$ where $k_{-}$, $k_{+}$ and $k^{*}$ are all positive. Using the fact $K(z)$ is positive allows one to re-parameterise the height according to
\begin{equation}\label{eq:change_of_variable}
 z\mapsto s=\int_{0}^{z}\frac{1}{K(\sigma)}d\sigma,
\end{equation}
where $s\in\left[0,\infty\right)$. Introducing the complex function
\begin{equation}\label{eq:ComplexVelocity}
 \Psi(s)=U(s)+iV(s),
\end{equation}
where $ U=u-u_{g}$ and $V=v-v_{g}$ are the components of the ageostrophic wind velocity in the zonal and meridional directions,
we may reformulate the system \eqref{eq:Ekman2Component} according to
\begin{equation}\label{eq:EkmanComplexForm}
 \frac{d^{2}\Psi}{ds^2}=i\alpha(s)\Psi(s)\quad \text{where }\alpha(s)=fK(z(s)).
\end{equation}
The boundary conditions \eqref{eq:PrandtlBC}--\eqref{eq:GeostrophicBC} become
\begin{equation}\label{eq:ComplexBC}
\begin{cases}
 \Psi=-u_{g}-iv_{g}\text{ at }s=0\\
 \Psi\to0\text{ as }s\to\infty.
 \end{cases}
\end{equation}
We note that since $\frac{ds}{dz}>0$, the monotone characteristics of $K(z)$ are preserved in $\alpha(s)$.

The general solution of the system \eqref{eq:EkmanComplexForm} may be written according to
\begin{equation}\label{eq:EkmanGeneralSolution}
 \Psi(s)=C_{-}\Psi_{-}(s)+C_{+}\Psi_{+}(s),
\end{equation}
where the basis solutions $\Psi_{\pm}(s)$ are defined according to their asymptotic behaviour
\begin{equation}\label{eq:Psi_asym}
 \Psi_{\pm}(s)\simeq e^{\mp\left(1+i\right)\lambda_{0}s}\text{ as } s\to \infty,
\end{equation}
where the conditions \eqref{eq:Psi_asym} are deduced from the asymptotic form of the system \eqref{eq:EkmanComplexForm} itself, namely
\begin{equation}\label{eq:EkmanComplexAsymptotic}
 \frac{d^{2}\Psi}{ds^2}\simeq ifk^{*}\Phi\text{ as }s\to\infty.
\end{equation}
The Wronskian of this basis is given by
\begin{equation}\label{eq:Wronskian}
    \mathbf{W}\left[\Psi_{+},\Psi_{-}\right]=\Psi_{+}\frac{d\Psi_{-}}{ds}-\frac{d\Psi_{+}}{ds}\Psi_{-}
\end{equation}
and satisfies
\begin{equation}
 \frac{d}{ds}\mathbf{W}\left[\Psi_{+},\Psi_{-}\right]=0,
\end{equation}
as a result of \eqref{eq:EkmanComplexForm}, in which case we may use the asymptotic form of $\Psi_{\pm}(s)$ to obtain
\begin{equation}
 \mathbf{W}\left[\Psi_{+},\Psi_{-}\right]=2(1+i)\lambda_{0}\neq 0.
\end{equation}
Hence, the Wronskian of the solution basis is non-zero, meaning this basis is linearly independent. The boundary condition $\Psi\to0$ as $s\to\infty$ ensures the physically relevant solution to the system \eqref{eq:EkmanComplexForm} is of the form
\begin{equation}\label{eq:ComplexPhysicalSolution}
 \Psi(s)=C_{+}\Psi_{+}(s),
\end{equation}
which remains stable as $s\to\infty$.

\subsection{The nonlinear formulation}
Reformulating the system \eqref{eq:EkmanComplexForm} using polar coordinates, namely
\[\Psi(s)=\rho(s)e^{i\tau(s)}\]
with
\begin{equation}
 U(s)=\rho(s)\cos(\tau(s))\quad V(s)=\rho(s)\sin(\tau(s)).
\end{equation}
a variety of solutions for the eddy viscosity $K(z)$ become accessible, cf. \cite{CJ2019}.
We now find $\Psi^{\prime\prime}=i\alpha(s)\Psi$ may be alternatively written in terms of real and imaginary parts, according to
\begin{subequations}
\begin{align}\label{nonlinODEa}
\rho^{\pprime}\cos(\tau)-2\rho^{\prime}M^{\prime}\sin(\tau)-\rho\rho^{\pprime}\sin(\tau)-\rho\rho^{\prime 2}\cos(\tau)&=-\alpha\rho\sin(\tau)\\
\label{nonlinODEb}
\rho^{\pprime}\sin(\tau)+2\rho^{\prime}\rho^{\prime}\cos(\tau)+\rho\rho^{\pprime}\cos(\tau)-\rho\rho^{\prime 2}\sin(\tau)&=\alpha \rho\cos(\tau).
\end{align}
\end{subequations}
We combine the above equations according to $\eqref{nonlinODEa}\times \cos(\tau) + \eqref{nonlinODEb}\times \sin(\tau)$ and $\eqref{nonlinODEa}\times (-\rho\sin(\tau)) + \eqref{nonlinODEb}\times (\rho\cos(\tau))$ to yield the following:
\begin{subequations}
\begin{align}\label{nlsysa}
 \rho^{\pprime}-\rho\tau^{\prime 2}&=0\\
 \label{nlsysb}
 (\rho^2\tau^{\prime})^{\prime}-\alpha \rho^2&=0,
\end{align}
\end{subequations}
where equations \eqref{nlsysa} \& \eqref{nlsysb} correspond to $\Psi^{\pprime}=i\alpha\Psi$ whenever $\rho\neq 0$. Introducing
\begin{equation}
 u_{g}+iv_{g}=\rho_{g}e^{i\tau_{g}},
\end{equation}
the boundary conditions \eqref{eq:ComplexBC} become
\begin{equation}
\begin{cases}
 \tau(\infty)=\tau_{g}\\
 \tau(0)=\tau_{g}+\pi.
\end{cases}
\end{equation}
As shown in \cite{CJ2019}, the modulus $\rho(s)$ is convex and satisfies
\begin{equation}\label{cond:M}
\begin{rcases}
 \rho(s)>0\\
 \rho^{\prime}(s)<0\\
\end{rcases} \text{ for }s>0,
\end{equation}
while the argument satisfies
\begin{equation}\label{cond:tau}
 \tau^{\prime}(s)<0\text{ for }s\geq 0.
\end{equation}
Using equation \eqref{nlsysa} to eliminate $\tau^{\prime}(s)$ yields
\begin{equation}\label{eq:tauprime}
\tau^{\prime}(s)=-\sqrt{\frac{\rho^{\pprime}(s)}{\rho(s)}},
\end{equation}
and substituting this expression for $\tau^{\prime}(s)$ into equation \eqref{nlsysb}, we infer
\begin{equation}\label{eq:alphaM}
 \alpha(s)=-\frac{3\rho^{\prime}\rho^{\pprime}+\rho\rho^{\prime\pprime}}{2\rho\sqrt{\rho\rho^{\pprime}}}.
\end{equation}
Thus we may obtain the eddy viscosity profile from the wind speed alone.

\section{A separable ODE for the ageostrophic wind speed}\label{sec:sepODE}
Given an appropriate ageostrophic wind speed $\rho(s)$, restricted by the conditions \eqref{cond:M} and the requirement that $\rho(s)$ be a convex function of $s$ (cf. equation \eqref{eq:tauprime}), we may reconstruct the eddy viscosity profile associated with this Ekman flow. In \cite{CJ2019} the authors choose explicit examples of weakly decaying and exponentially decaying wind speed profiles and outline the process of obtaining the deflection angle and associated eddy diffusion coefficient associated with these flows.

In this paper we wish to analyse a general class of convex, exponentially decaying ageostrophic wind speeds of the form
\begin{equation}\label{rho:exp}
\rho(s)=e^{\mu(s)}\text{ for }0\leq s < \infty,
\end{equation}
subject to the conditions \eqref{cond:M} and $\alpha(s)>0$. We note that the condition $\rho^{\prime}(s)<0$ also requires $\mu^{\prime}(s)<0$ for $s>0$, while convexity of $\rho(s)$ means $\rho^{\pprime}(s)>0$, which written in terms of $\mu(s)$ becomes
\begin{equation}\label{cond:muprime}
\left(\mu^{\prime}(s)^2+\mu^{\pprime}(s)\right)e^{\mu(s)}>0.
\end{equation}
We propose a separable ordinary differential for $\mu^{\prime}(s)$ of the form
\begin{equation}\label{eq:ODE_separable}
 \mu^{\pprime}+\mu^{\prime2}=\gamma(s)\mu^{\prime2}.
\end{equation}
where $\gamma(s)>0 \text{ for }s>0,$ thus ensuring \eqref{cond:muprime} is preserved. Moreover, since
\[\frac{\rho^{\pprime}}{\rho}=\mu^{\prime 2}+\mu^{\pprime},\]
we find that equation \eqref{eq:tauprime} may be reformulated according to
\begin{equation}\label{eq:tauprime_separableode}
\tau^{\prime}(s)=\mu^{\prime}(s)\sqrt{\gamma(s)},
\end{equation}
which agrees with condition \eqref{cond:tau} when $\mu^{\prime}(s)<0$ and $\gamma(s)>0$ for $s>0$.
We observe that for the general class of wind speed profiles governed by equation \eqref{eq:ODE_separable}, the relationship between $\alpha(s)$ and $\rho(s)$ given by equation \eqref{eq:alphaM} may be reformulated as
\begin{equation}\label{eq:alphamu}
 \alpha(s)=-\frac{d}{ds}\sqrt{\mu^{\prime2}+\mu^{\pprime}}-2\mu^{\prime}\sqrt{\mu^{\prime2}+\mu^{\pprime}}=-\left(\frac{d}{ds}+2\mu^{\prime}\right)\sqrt{\gamma(s)\mu^{\prime 2}},
\end{equation}
which will prove useful in what follows.


\subsection{The slowly decaying solution}\label{sec:slow}
Obviously the simplest case to analyse is of the form
\[\gamma(s)=1+a>0,\]
where $a$ is constant. It is straight forward to show that
\begin{equation}\label{eg1:muprime}
\mu^{\prime}(s)=-\frac{b}{1+abs},
\end{equation}
where we introduce the integration constant $\mu^{\prime}(0)=-b<0$. Given a sufficiently large $s$ we observe that
\begin{equation}
 \mu^{\prime}(s)\simeq-\frac{1}{as}<0,
\end{equation}
in which case the condition $\mu^{\prime}(s)<0$ for all $s>0$ requires $a>0$. Integrating equation \eqref{eg1:muprime}, we  find
\begin{equation}
 \mu(s)=\mu_{g}+\ln(1+abs)^{-\frac{1}{a}},
\end{equation}
with  $\mu_{g}$ defined according to $\rho_{g}=e^{\mu_{g}}$, where $\rho_{g}$ is the geostrophic wind speed.  Hence the speed profile is given by
\begin{equation}\label{eg1:M}
\rho(s)=\rho_{g}\left(1+abs\right)^{-\frac{1}{a}}.
\end{equation}
while the associated deflection angle is given by
\begin{equation}\label{eg1:tau}
\begin{aligned}
&\tau^{\prime}(s)=-\frac{b\sqrt{1+a}}{1+abs} \Rightarrow \tau(s)=\tau_{g}+\pi-\frac{\sqrt{1+a}}{a}\ln\left(1+abs\right)
\end{aligned}
\end{equation}
with $\tau_g$ the direction of the geostrophic wind at the top of the Ekman layer. Equations \eqref{eq:alphamu}--\eqref{eg1:muprime} yield an eddy viscosity  coefficient given by
\begin{equation}\label{eg1:alpha}
 \alpha(s)=\frac{\alpha_{0}}{(1+abs)^{2}},
\end{equation}
and with  $\alpha_{0}>0$ it follows that $\alpha(s)>0$ for all $s>0$, in line with the conditions for $K(z)$ proposed in \cite{CJ2019}.

\subsubsection{The slowly decaying solution in physical variables}\label{sec:slow_physical}
It follows from equation \eqref{eq:change_of_variable} that $\frac{d}{ds}=K\frac{d}{dz}$, and using the notation $F^{\prime}=\frac{dF}{ds}$ and $\dot{F}=\frac{dF}{dz}$ for any function $F$, we may interpret the coefficients $a$ and $b$ in terms of $\rho_{g}=\rho(0)$, $K_{0}=K(0)$ and the vertical gradients $\dot{\rho}_{0}=\dot{\rho}(0)$ and $\dot{\tau}_{0}=\dot{\tau}(0)$. Using this notation, we evaluate equations \eqref{eg1:muprime} and \eqref{eg1:tau} at $s=z=0$ to yield
\begin{equation}\label{eg1:b}
b=-\frac{K_{0}\dot{\rho}_{0}}{\rho_{g}} ,\quad -b\sqrt{1+a}=K_{0}\dot{\tau}_{0},
\end{equation}
and since $b$, $K_{0}$ and $\rho_{g}$ are all positive, it indicates the ageostrophic wind speed decreases with height near the bottom, as expected. Equation \eqref{eg1:b} yields
\begin{equation}\label{eg1:a}
a=\omega^2-1,\quad \omega\equiv\frac{\rho_{g}\dot{\tau}_{0}}{\dot{\rho}_0}
\end{equation}
where this definition of $\omega$ will be used throughout. Since we require $a>0$ we must impose $\omega^2>1$ for the weakly decaying model. Lastly, we also note from equations \eqref{eq:EkmanComplexForm} and \eqref{eg1:alpha} that $\alpha_{0}=fK_{0}$.

To re-write the expressions for $\rho$, $\tau$ and $K$ in terms of the vertical coordinate $z$, we observe from equation \eqref{eq:change_of_variable} that $dz=\frac{\alpha(s)}{f}ds$. Using equation \eqref{eg1:alpha} we may integrate explicitly to find $z(s)$, thereby allowing us to deduce its inverse
\begin{equation}\label{eg1:sz}
s(z)=\frac{z}{K_{0}-abz}=\frac{z}{K_{0}\left(1-\frac{z}{h}\right)},
\end{equation}
where we introduce the height-parameter $h=-\frac{\rho_{g}}{\dot{\rho}_0\left(\omega^2-1\right)}>0$.

Equations \eqref{eg1:M} and \eqref{eg1:b}--\eqref{eg1:sz} allow us to write
\begin{equation}\label{eg1:M(z)}
\rho(z)=\begin{cases}
 \rho_{g}\left(1-\frac{z}{h}\right)^{\frac{1}{\omega^2-1}}\text{ for }z\in[0,h)\\
 0 \text{ for }z\in[h,\infty).
\end{cases}
\end{equation}
The corresponding deflection angle is of the form
\begin{equation}\label{eq:tau(z)}
\tau(z)=
\begin{cases}
\tau_{g}+\pi+\frac{\omega}{\omega^2-1}\ln\left(1-\frac{z}{h}\right),\text{ for }z\in[0,h)\\
\tau_{g}\text{ for }z\in[h,\infty)
\end{cases}
\end{equation}
which is a monotonically decreasing function of $z$, meaning the ageostrophic wind direction rotates  clockwise  with increasing height, as expected. Moreover, the associated eddy viscosity profile is simply given by
\begin{equation}\label{eg1:Kz}
K(z)=
\begin{cases}
 K_{0}\left(1-\frac{z}{h}\right)^{2},\text{ for }z\in[0,h)\\
 0\text{ for }z\in[h,\infty),
\end{cases}
\end{equation}
and so clearly we have $K(z)\to0$ and $\frac{dK}{dz}\to0$ as $z\to h$, properties shared with eddy viscosity profiles previously investigated in \cite{OBr1970} for example.

\subsubsection{The relative deflection angle \& the height of the Ekman layer}

The relative deflection of the the flow is the angle between the wind direction at height $s$ and the direction of the geostrophic wind and is given by
\begin{equation}\label{eq:rel_del}
\beta(s)=\arctan\left(\frac{\rho(s)\sin(\tau(s)-\tau_{g})}{\rho(s)\cos(\tau(s)-\tau_{g})+\rho_{g}}\right),
\end{equation}
cf. \cite{CJ2019}. This deflection angle may be calculated at the base of the Ekman layer $s=0$ using l'H\^{o}pital's rule to give
\begin{equation}\label{eq:beta_max}
\beta(0)=\arctan\left(\frac{\rho_{g}\tau^{\prime}(0)}{\rho^{\prime}(0)}\right)=\arctan\left(\frac{\rho_{g}\dot{\tau}_{0}}{\dot{\rho}_{0}}\right)=\arctan(\omega).
\end{equation}
Thus we see the significance of the parameter $\omega$, it is the tangent of the relative angle between the wind direction at the bottom of the PBL and the geostrophic wind at the top. Given that we require $\omega>1$ for the weakly decaying case (cf. equation \eqref{eg1:a}), it follows this model is only appropriate when the angle between the wind directions at the bottom and top of the atmospheric boundary layer exceeds $45^\circ$. Such scenarios are known to arise, for instance field data from the Tibetan plateau reveal relative deflections above $50^{\circ}$ (see \cite{ZXW2003}).

A useful definition for the height of the Ekman layer is as the smallest value $z=\he$ where the wind-direction is aligned with the geostrophic wind-direction, cf. \cite{CJ2019}. It follows that $\tau(\he)=\tau_{g}$ and equation \eqref{eq:tau(z)} means this height $\he$ is explicitly given by
\begin{equation}
  \he=h\left(1-e^{-\frac{\pi}{\omega}\left(\omega^2-1\right)}\right).
\end{equation}
A useful feature of this definition for the height of the Ekman layer is that it may be obtained from ground based measurements of the flow and the geostrophic wind speed, which is essentially constant above this height.

\begin{figure}[ht!]
\centering
\includegraphics[width=0.33\textwidth]{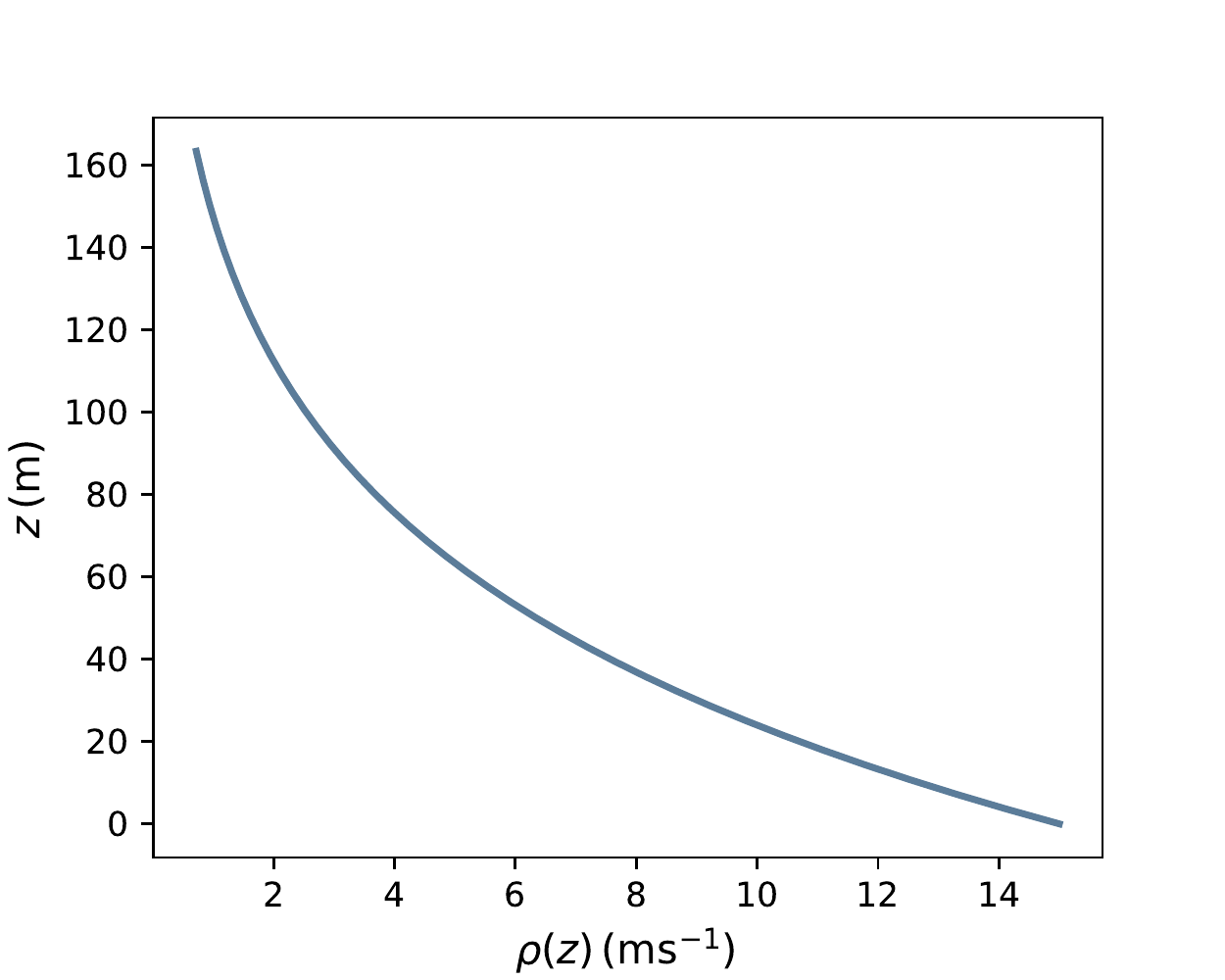}\includegraphics[width=0.33\textwidth]{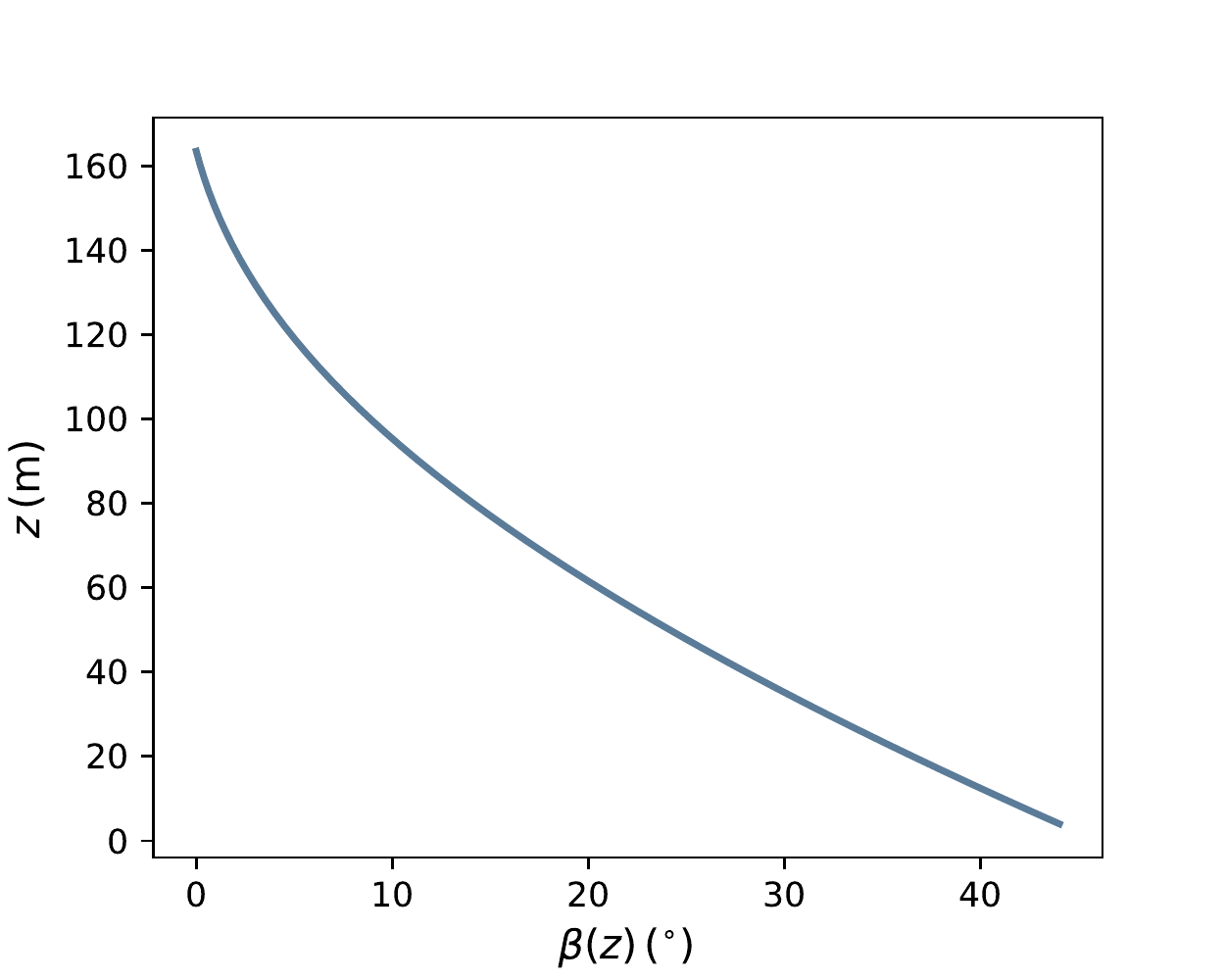}\includegraphics[width=0.33\textwidth]{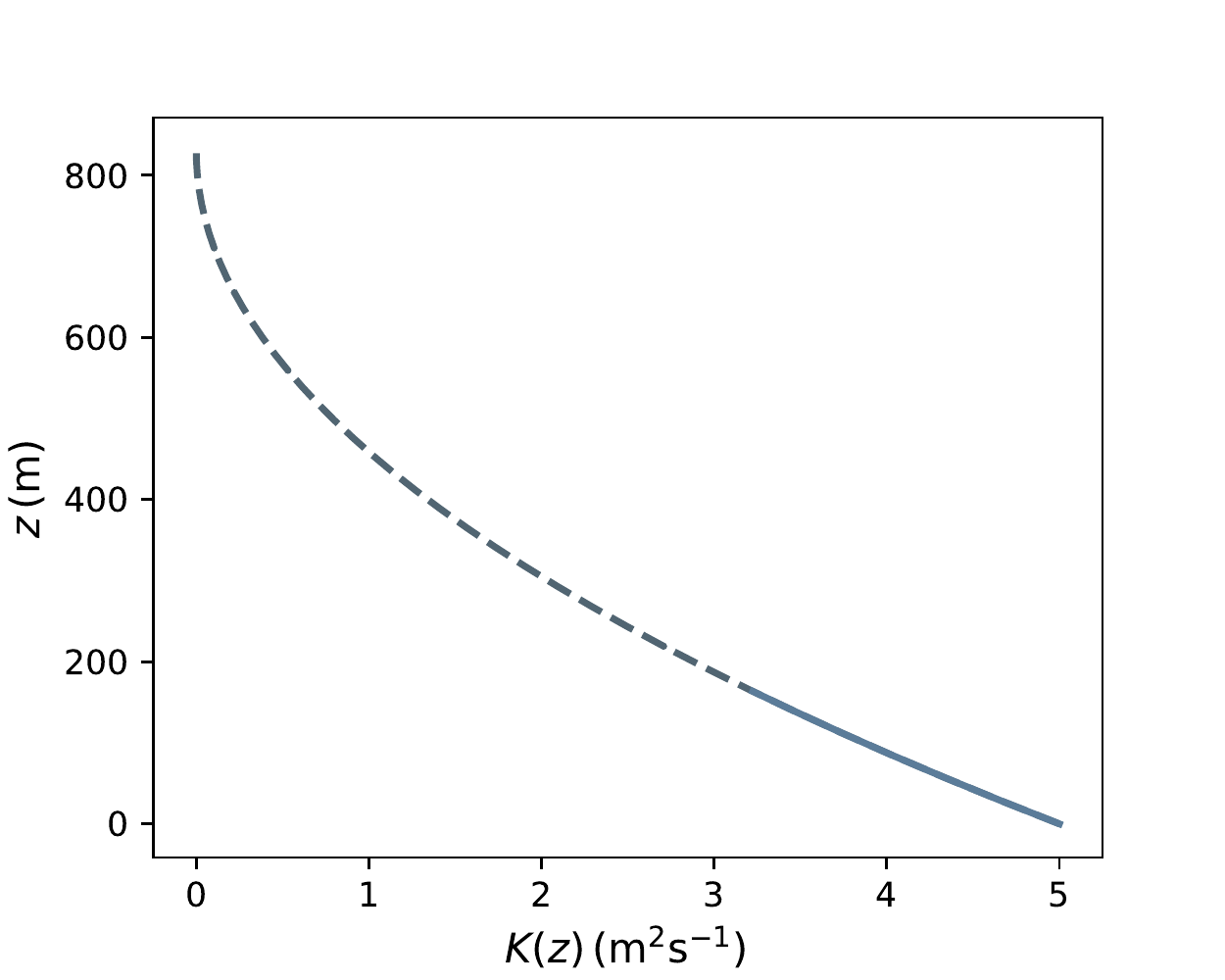}
\includegraphics[width=\textwidth]{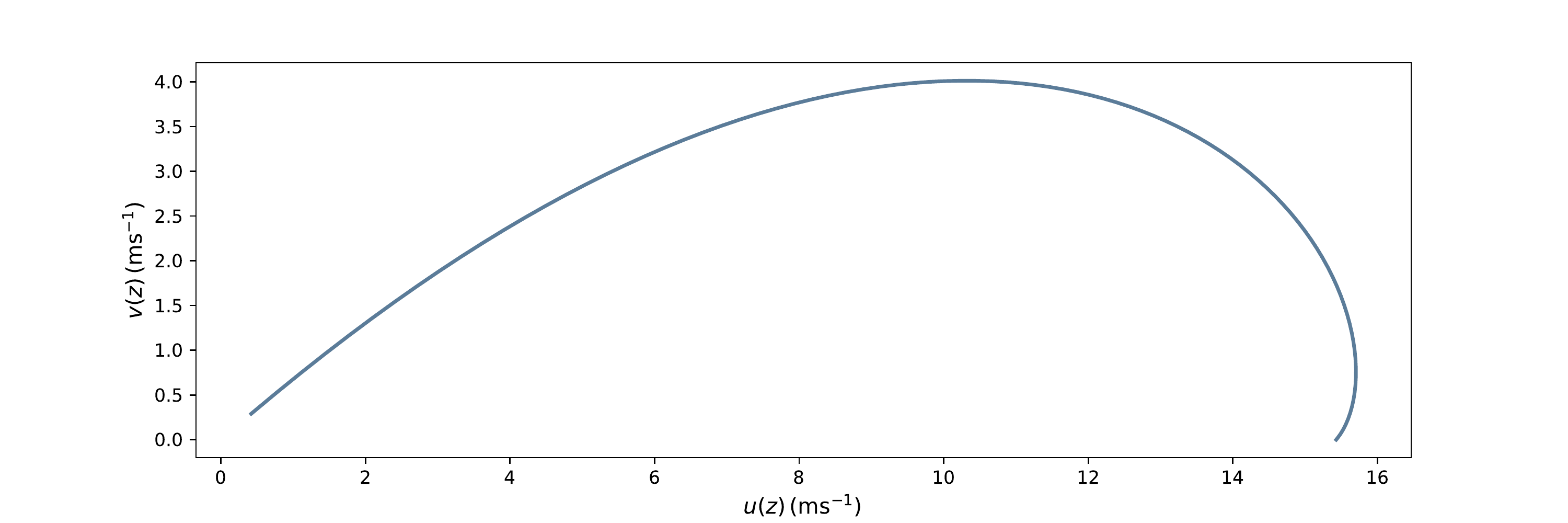}
\caption{ The wind speed $\rho(z)$, relative deflection $\beta(z)$ and eddy viscosity $K(z)$ as well as the hodograph of the \textbf{slowly decaying model}.  The parameters used in this model are $\rho_{g}=15\,\mathrm{ms^{-1}}$, $\dot{\rho}_{0}=-0.25\,\mathrm{s^{-1}}$ and a maximum relative deflection $\beta_{0}=46^{\circ}$. The eddy viscosity coefficient at the bottom of the PBL is $K_{0}=5\mathrm{m^2s^{-1}}$. The profile of the eddy viscosity coefficient shows the decay over the Ekman layer $0<z\lesssim163\,\mathrm{m}$ (solid line). }\label{fig1}
\end{figure}

We note that $K(z)\to0$ as $z\to h$, which appears to contradict the condition $K:[0,\infty)\to[k_{-},k_{+}]$ with $k_{\pm}$ positive constants. However, as applied in \cite{CJ2019} it appears this condition is a sufficient condition for the existence of a solution of \eqref{eq:EkmanComplexForm}. Furthermore, in  \cite{PKB2005} the authors develop analytic solutions for atmospheric Ekman flows with slowly varying eddy viscosity profiles where $K(z)\geq0$, so a vanishing eddy viscosity appears to be physically reasonable also.
In \cite{CJ2019} the authors proposed an ab-initio ageostrophic wind speed profile of the form
\begin{equation}\label{eq:CJMs1}
 \rho(s)=\begin{cases}
          \frac{\tilde{b}}{1+\tilde{a}s}, \text{ for }0\leq s\leq s_{0}\\
          \frac{\tilde{b}}{1+\tilde{a}s_{0}}, \text{ for }s>s_{0},
         \end{cases}
\end{equation}
with $\tilde{a}$, $\tilde{b}$ and $s_{0}$ all positive constants. Thus we see that \eqref{eg1:M} and \eqref{eq:CJMs1} to a large extent are the same wind speed profile when
\begin{equation}\label{eg1:Ly_to_CJ}
a=1,\quad b=\tilde{a},\quad \rho_{g}=\tilde{b},\quad s_{0}=s(\he)=\frac{\he}{K_{0}\left(1-\frac{\he}{h}\right)}
\end{equation}
The wind speed profile \eqref{eg1:M} is a generalisation of the profile \eqref{eq:CJMs1} in the sense that the decay rate of the profile \eqref{eg1:M} may be altered by varying the value of the parameter $a$ (or equivalently the physical parameter $\omega$). On the other hand the eddy viscosity profiles $K(z)$ associated with the speed profiles \eqref{eg1:M} and \eqref{eq:CJMs1} are basically the same when we impose \eqref{eg1:Ly_to_CJ}, with both profiles decaying quadratically as $z$ increases.  In figure \ref{fig1} the graphs of the wind speed $\rho(z)$, the  relative deflection $\beta(z)$ and the eddy diffusion $K(z)$ are shown in the top three panels, while the lower panel is the hodograph of the wind velocity (the graph of $v(z)$ vs. $u(z)$ for $z\in[0,\he)$) for the slowly decaying model.

\subsection{The exponentially decaying solution}\label{sec:exponential}
In this case we consider the ageostrophic wind speed governed by the separable ODE
\begin{equation}\label{eg2:ode}
 \mu^{\pprime}+\mu^{\prime2}=\left(1-\frac{\lambda^2}{(as+b)^2}\right)\mu^{\prime2} \text{ for }s>0, \text{ with } \mu_{0}^{\prime}=-\frac{ab}{\lambda},
\end{equation}
where $a$, $b$ and $\lambda$ are positive constants, while the condition $b^2\geq \lambda^2$  ensures $\gamma(s)=1-\frac{\lambda^2}{(as+b)^2}>0$ for all $s>0$. Separation of variables allows us to integrate to obtain
\begin{equation}\label{eg2:Ms}
\begin{aligned}
 \rho(s)=\rho_{g}e^{-\frac{a^2 s^2}{2\lambda^2}-\frac{a b}{\lambda^2}s},
\end{aligned}
\end{equation}
in which case the profile investigated in \cite{CJ2019} is reproduced when $\lambda=\sqrt{2}$ and $b=2$. Equations \eqref{eq:tauprime_separableode} and  \eqref{eg2:Ms} yield
\begin{equation}\label{eg2:taus}
\tau^{\prime}(s)=-\frac{a}{\lambda^2}\sqrt{(as+b)^2-\lambda^2},
\end{equation}
whose integral we will compute later.

We apply equation \eqref{eq:alphaM} to obtain the eddy viscosity coefficient
\begin{equation}\label{eg2:alpha}
 \alpha(s)=\frac{a^2(as+b)\left[2(as+b)^2-3\lambda^2\right]}{2\lambda^4\sqrt{(as+b)^2-\lambda^2}}.
\end{equation}
and to ensure $\alpha(s)$ is positive for all $s>0$ it is clear that we actually require $b^2\geq\frac{3}{2}\lambda^2$, which also ensures $\alpha(s)$ is bounded for all $s\geq0$. Integrating with respect to $s$, we have
\begin{equation}\label{eg2:z}
 z(s)=\frac{a}{6f\lambda^4}\left[\sqrt{x^2-\lambda^2}\left(2x^2-3\lambda^2\right)\right]_{x=b}^{x=as+b}.
\end{equation}
Since  $\alpha(s)>0\Rightarrow z^{\prime}(s)>0$ for $s>0$ when $b^2>\frac{3}{2}\lambda^2$ with $z(s)$ also being a continuous function of $s$ for all $s\in\left[0,\infty\right)$, an inverse function $s(z)$ for all $z\geq0$ is ensured, under an appropriate choice of parameters $a$, $b$ and $\lambda$ cf. \cite{Die1960}.

\subsubsection{Exponential decay in physical variables}
Applying the notation from section \ref{sec:slow_physical} to equations \eqref{eg2:Ms} and \eqref{eg2:taus}, we evaluate $\rho^{\prime}(0)$ and $\tau^{\prime}(0)$, to find
\begin{equation}\label{eg2:ics}
-\frac{ab}{\lambda^2}=\frac{K_{0}\dot{\rho}_{0}}{\rho_{g}},\quad
-a\frac{\sqrt{b^2-\lambda^2}}{\lambda^2}=K_{0}\dot{\rho}_{0},
\end{equation}
Multiplying and dividing these expressions separately, it may be deduced that
\begin{equation}\label{eg2:abl}
\begin{aligned}
 \frac{a^2}{\lambda^2}&=\frac{\K^2\dr^2(1-\omega^2)}{\rg^2}\equiv \w^2, \quad
 \frac{b^2}{\lambda^2}&=\frac{1}{1-\omega^2},
\end{aligned}
\end{equation}
where $\omega=\frac{\rg\dt}{\dr}$. The velocity scale $\w$ is interpreted as the rate at which horizontal momentum is transferred vertically near the bottom of the flow at $z=0$.  The condition $\alpha(0)=fK_{0}$ combined with these relations yields
\begin{equation}\label{eg2:w0}
   \w^2=\frac{2f\K\omega(1-\omega^2)}{\left(3\omega^2-1\right)},
\end{equation}
which relates this vertical velocity scale to latitude via the Coriolis parameter. The conditions $3\omega^2-1>0$ and $1-\omega^2>0$ ensure the model is only valid when the relative deflection angle between the geostrophic wind direction and the  wind at the bottom of the PBL is constrained by $30^{\circ}<\beta(0)<45^{\circ}$, which agrees with available field data (cf. \cite{Pena_etal2016,Rys_etal2016}).

\subsubsection{The inverse map}
To reformulate equation \eqref{eg2:z} in physical variables we introduce the notation
\begin{equation}\label{eg2:xi}
 \sqrt{(as+b)^2-\lambda^2}=a\sqrt{(s+s_1)^2-s_{0}^2} \equiv a\xi(s)
\end{equation}
where we define $s_{0}\equiv\frac{1}{w_{0}}$ and $s_{1}\equiv\frac{1}{\w\sqrt{1-\omega^2}}$ and
\begin{equation}\label{eg2:z0}
 \frac{a}{6f\lambda^2}\sqrt{b^2-\lambda^2}\left[2b^2-5\lambda^2\right]=\frac{\w\omega\left(3\omega^2+2\right)}{6f\sqrt{1-\omega^2}}\equiv{z_0}.
\end{equation}
Hence, we may recast equation \eqref{eg2:z} in the form of a cubic polynomial
\begin{equation}
 \xi^{3}-\frac{3s_{1}^2}{2}\xi-\frac{\eta(z)}{2}=0,\quad \eta(z)=\frac{5f}{\w^4}(z-z_{0}),
\end{equation}
whose only real root is given by
\begin{equation}
 \hat{\xi}(z)=\frac{\sqrt[3]{2\eta+\sqrt{\eta^2-2s_{1}^6}}+\sqrt[3]{2\eta-\sqrt{\eta^2-2s_{1}^6}}}{2}.
\end{equation}
The function $\hat{\xi}(z)$ is   the function $\xi(s)$, parameterised with respect to $z$, as opposed to $s$. Transposing equation \eqref{eg2:xi}, and using $\xi(s)\equiv\hat{\xi}(z)$, can write the inverse of $z(s)$ as follows:
\begin{equation}\label{eg2:sz}
s(z)=\sqrt{\hat{\xi}(z)^2+s_{0}^2}-s_{1}.
\end{equation}

\subsubsection{The height of the boundary layer}
\begin{figure}[ht!]
\centering
\includegraphics[width=0.333\textwidth]{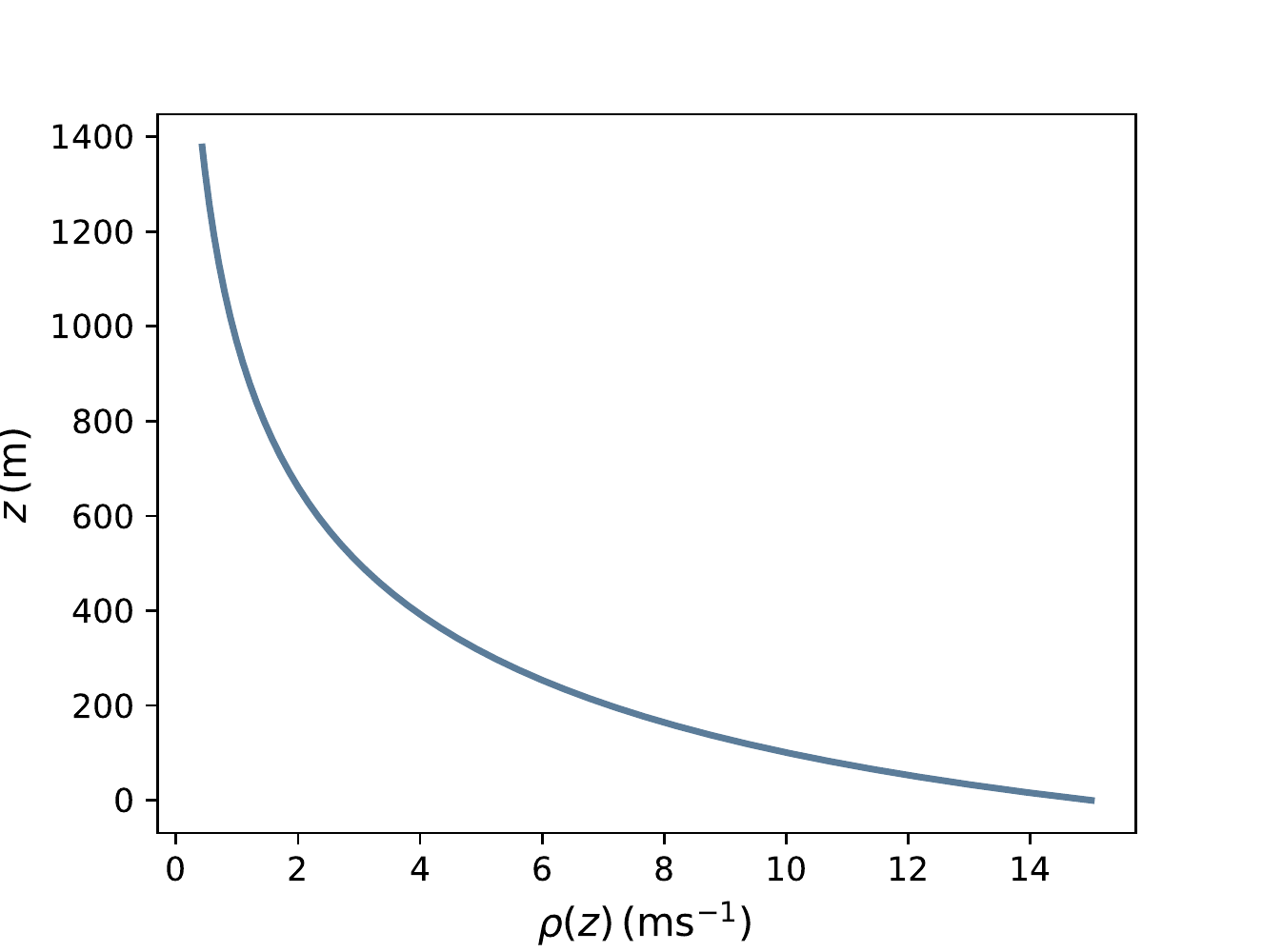}\includegraphics[width=0.333\textwidth]{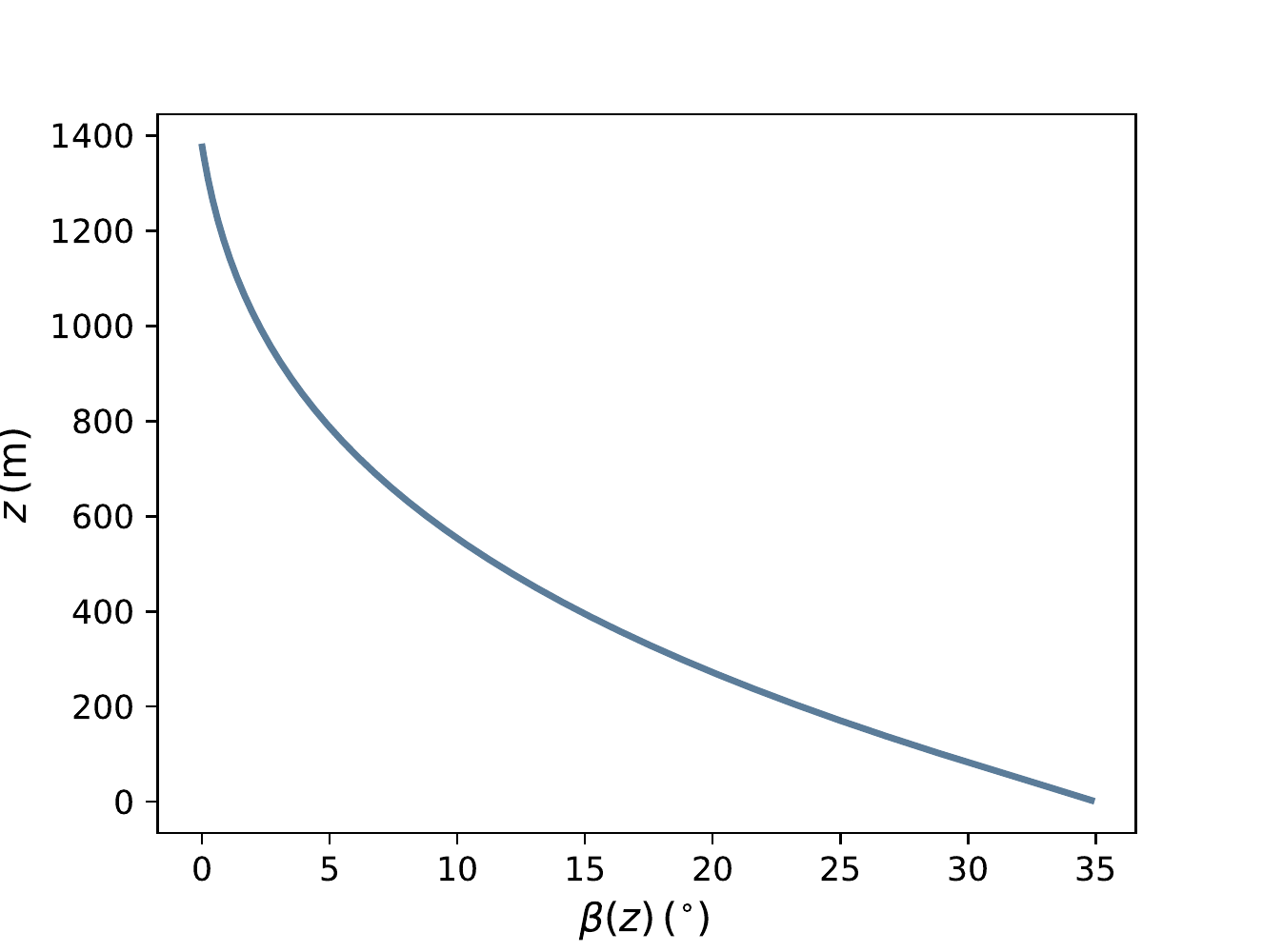}\includegraphics[width=0.333\textwidth,keepaspectratio]{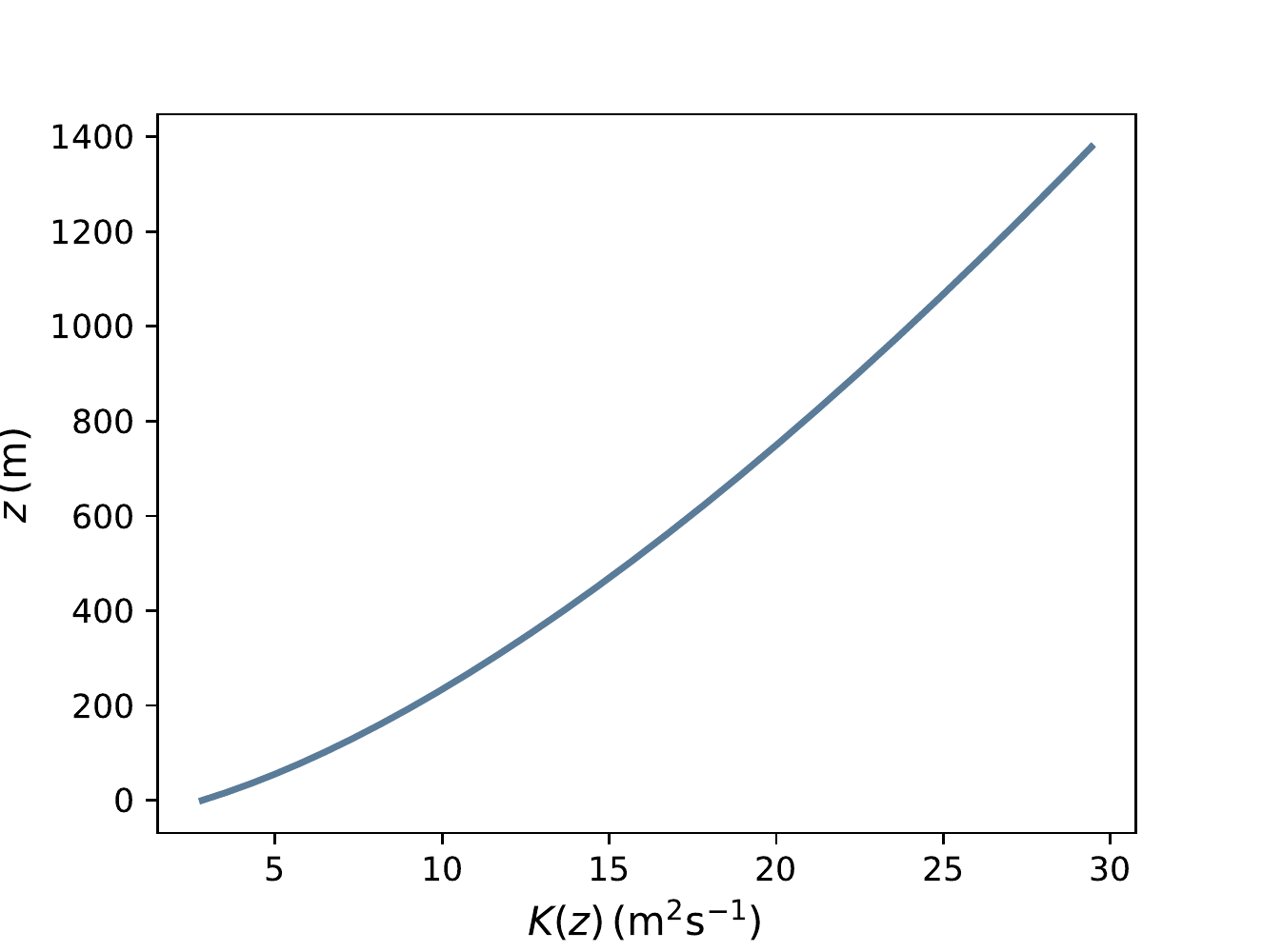}
\includegraphics[width=0.999\textwidth]{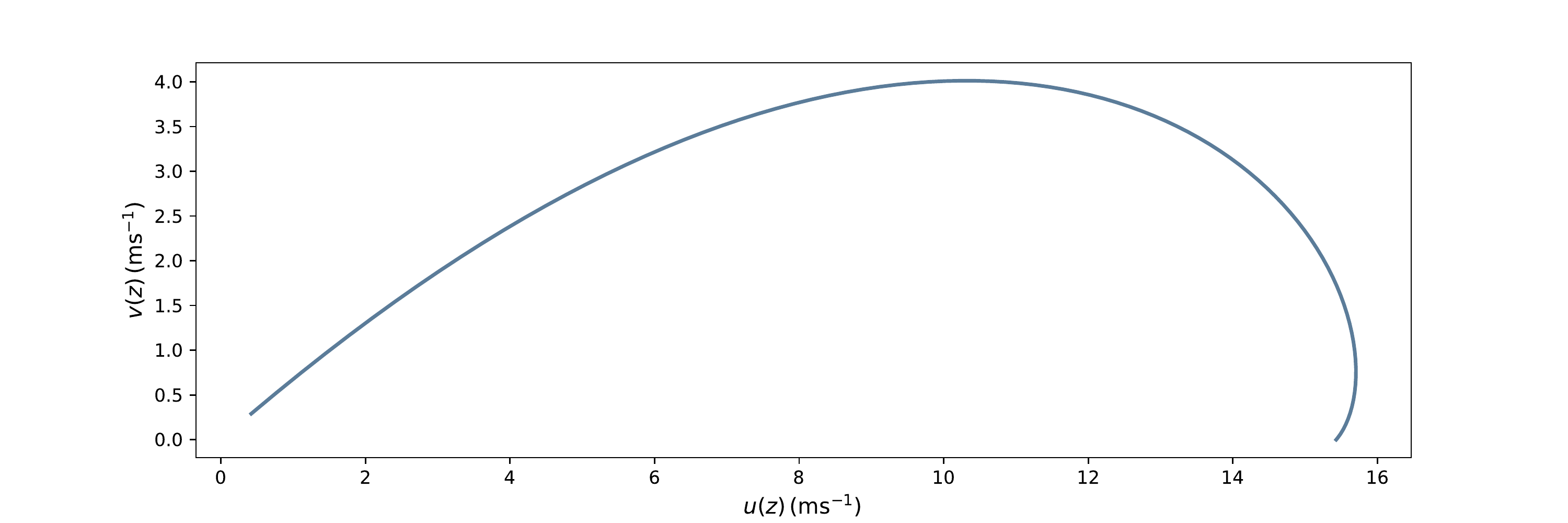}
\caption{ The wind speed $\rho(z)$, relative deflection $\beta(z)$ and eddy viscosity $K(z)$ as well as the hodograph of the \textbf{exponentially decaying model}. These profiles correspond to the choice of parameters $\rho_{g}=15\,\mathrm{ms^{-1}}$, $\w=0.02\,\mathrm{ms^{-1}}$ and a maximum relative deflection $\beta_{0}=35^{\circ}$ and at a latitude of $52^{\circ}$ north. Under this choice of parameters, the eddy viscosity coefficient at the bottom of the PBL is  $K_{0}=2.82\,\mathrm{m^2s^{-1}}$ and the Ekman layer has height $\he\simeq1379\,\mathrm{m}$.}\label{fig:exp_decay}
\end{figure}

Written in terms of the parameter $s$, we find that the ageostrophic wind speed may be written as
\begin{equation}
 \rho(s)=\rho_{g}\exp\left[-\frac{\w^2s^2}{2}-\frac{\w s}{\sqrt{1-\omega^2}}\right],
\end{equation}
Given the restriction $\frac{1}{3}<\omega^2<1$, it is clear that $\rho(s)$ is convex for all $s>0$, as expected.
The height of the boundary layer is defined as the smallest value $s=\se$ such that $\tau(\se)=\tau_{g}$, and so equation \eqref{eg2:taus} yields
\begin{equation}
\pi+\frac{1}{2}\ln\left[\sqrt{x^2-1}+x\right]_{x=\frac{s_1}{s_{0}}}^{\frac{\se+s_0}{s_{1}}}-\frac{1}{2}\left[x\sqrt{x^2-1}\right]_{x=\frac{s_1}{s_{0}}}^{\frac{\se+s_0}{s_{1}}}=0.
\end{equation}
An explicit expression for $\se$ in terms of $s_{0}$, $s_{1}$ is obviously not available from this implicit definition, however a numerical value is always assured for appropriate values of $s_{0}$ and $s_{1}$ as a consequence of the implicit function theorem. The ageostrophic wind speed $\rho$, the relative deflection $\beta$ and the eddy viscosity $K(z)$ for the exponentially decaying model are shown in the top three panels of figure \ref{fig:exp_decay}, while the lower panel of this figure shows the hodograph of the ageostrophic wind velocity in the PBL.

\subsection{Slow exponential decay}\label{sec:slow_exponential}
In this example we consider the separable ODE give by
\begin{equation}\label{eg3:ode}
 \mu^{\pprime}+\mu^{\prime2}=\left(1-e^{-(as+b)}\right)\mu^{\prime2},\quad \mu^{\prime}_{0}=-\frac{a}{e^{-b}+1},
\end{equation}
where $a$ and $b$ are positive constants. Integrating we find
\begin{equation}\label{eg3:Ms}
 \rho(s)=\rho_{g}\left(\frac{1+e^{b}}{1+e^{as+b}}\right).
\end{equation}
while the associated deflection angle is given by
\begin{equation}\label{eg3:rhos}
\tau(s)=\tau_g+\pi+\ln\left\vert\frac{\left(\sqrt{1-x}-1\right)\left(\sqrt{1-x}+\sqrt{2}\right)^{\sqrt{2}}}{\left(\sqrt{1-x}+1\right)\left(\sqrt{1-x}-\sqrt{2}\right)^{\sqrt{2}}}\right\vert_{x=e^{-b}.}^{x=e^{-(as+b)}}
\end{equation}
The eddy viscosity coefficient $\alpha(s)$ is given by
\begin{equation}\label{eg3:alpha}
 \alpha(s)=\frac{a^2e^{as+b}\left(4e^{2(as+b)}-7e^{as+b}+1\right)}{2\left(1+e^{as+b}\right)^2\sqrt{e^{2(as+b)}-e^{as+b}}},
\end{equation}
and using $dz=\frac{\alpha(s)}{f}ds$ we obtain after integration
\begin{equation}\label{eg3:zs}
 z(s)=\left[\sqrt{2}\ln\left(\frac{\sqrt{1-x}+\sqrt{2}}{\sqrt{1-x}-\sqrt{2}}\right)+\ln\left(\frac{\sqrt{1-x}-1}{\sqrt{1-x}+1}\right)\right]_{x=e^{-b}.}^{x=e^{-(as+b)}}
\end{equation}
Since we require $\alpha(s)>0$ for all $s>0$ and the polynomial $4\xi^2-7\xi+1=0$ has roots at $\frac{7\pm{\sqrt{33}}}{8}$, we infer from equation \eqref{eg3:alpha} that $b$ must satisfy the inequality
\begin{equation}\label{eg3:bcondition}
b\geq\ln\left(\frac{7+\sqrt{33}}{8}\right),
\end{equation}
This in turn ensures $z(s)$ is a monotonic function of $s$ and therefore invertible in principle. As an exact expression for the inverse of this function cannot be found we calculate the inverse numerically once an appropriate choice is made for the parameters $a$ and $b$.

\subsubsection{Physical parameters of slow exponential decay}
Using the notation from section \ref{sec:slow_physical}, we find that the parameters $a$ and $b$ are related to the physical parameters $K_{0}$, $\dot{\rho}_{0}$, $\dot{\tau}_{0}$ and $\rho_{g}$ by the conditions
\begin{equation}\label{eg3:parameters}
 -\frac{a}{1+e^{-b}}=\frac{K_{0}\dot{\rho}_{0}}{\rho_{g}},~ -\frac{a\sqrt{1-e^{-b}}}{1+e^{-b}}=K_{0}\dot{\tau}_{0},
\end{equation}
which may be solved to yield
\begin{equation}\label{eq3:ab}
 \frac{a^2\sqrt{1-e^{-b}}}{\left(1+e^{-b}\right)^2}=\w^2,~
 e^{-b}=1-\omega^2,
\end{equation}
where $\omega=\frac{\rg\dt}{\dr}$ and $\w^2=\frac{\K^2\dr\dt(2-\omega^2)}{\rg\omega^2}$. The condition $\alpha(0)=f\K$ now gives
\begin{equation}\label{eg3:k0}
 \w^2=\frac{2fK_0\omega(2-\omega^2)}{\left(\omega^4+5\omega^2-2\right)},
\end{equation}
where again we interpret $w_{0}$ as the rate at which horizontal momentum is transported in the vertical direction due to the effects of eddy viscosity.

\begin{figure}[ht!]
\centering
\includegraphics[width=0.333\textwidth]{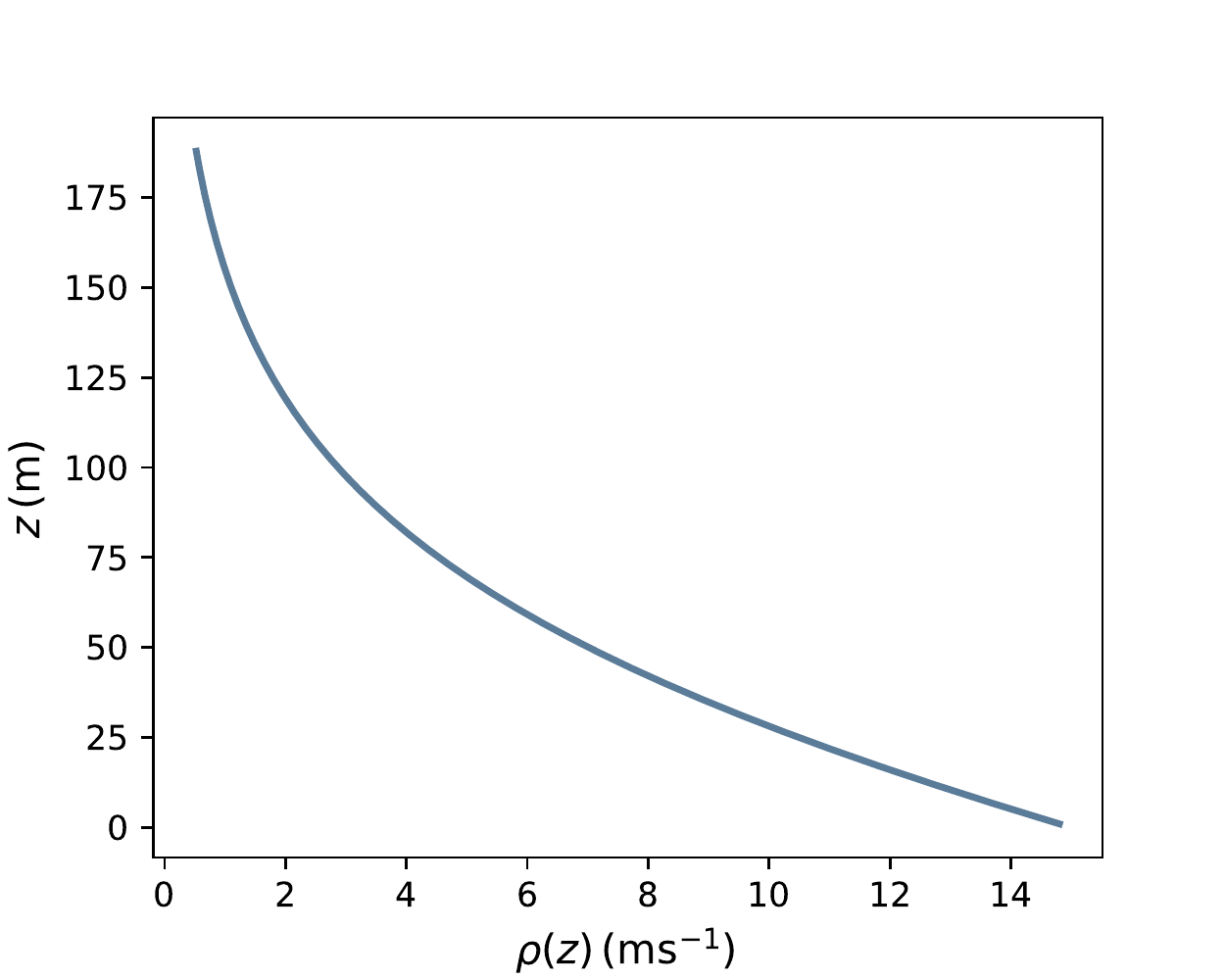}\includegraphics[width=0.333\textwidth]{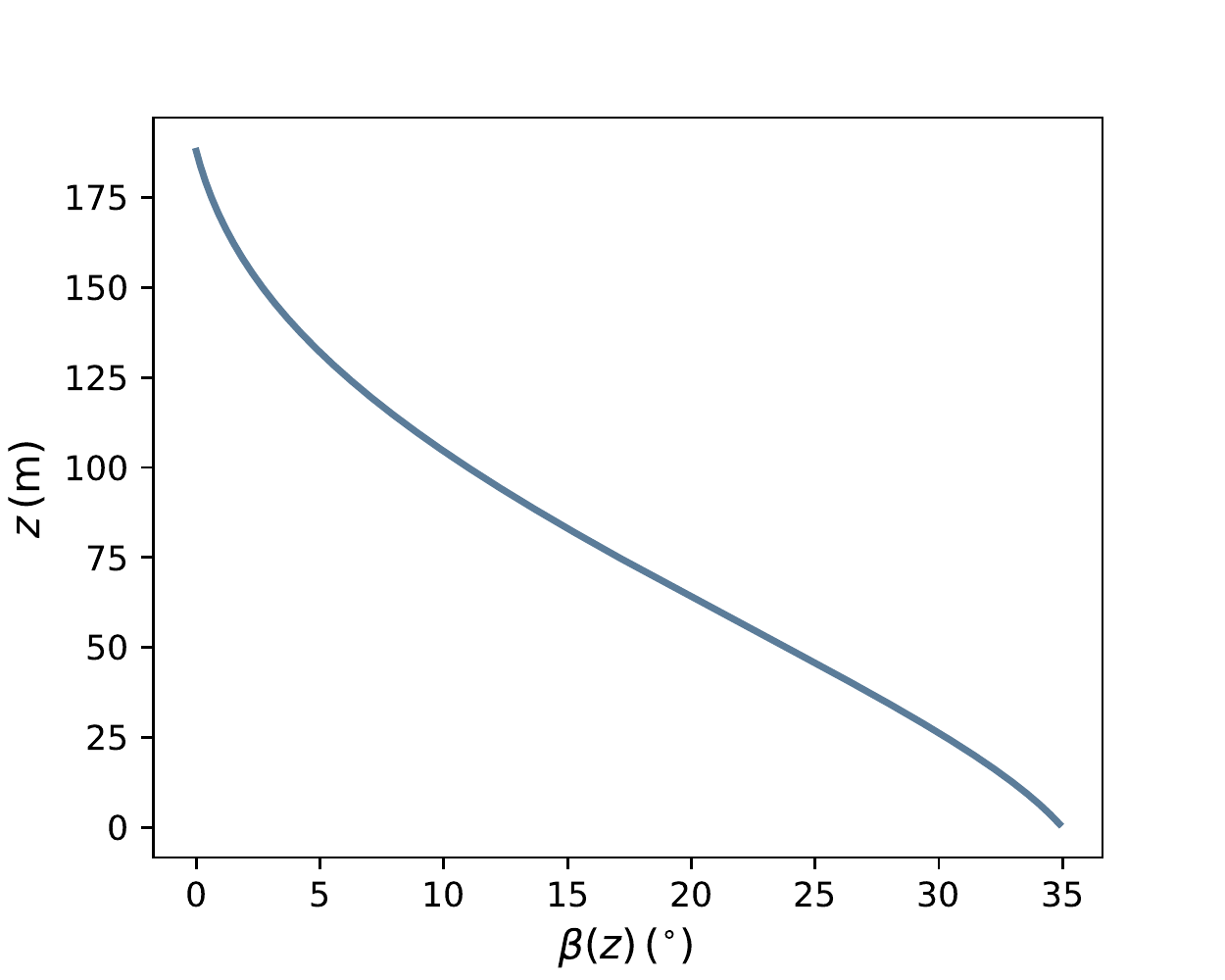}\includegraphics[width=0.333\textwidth,keepaspectratio]{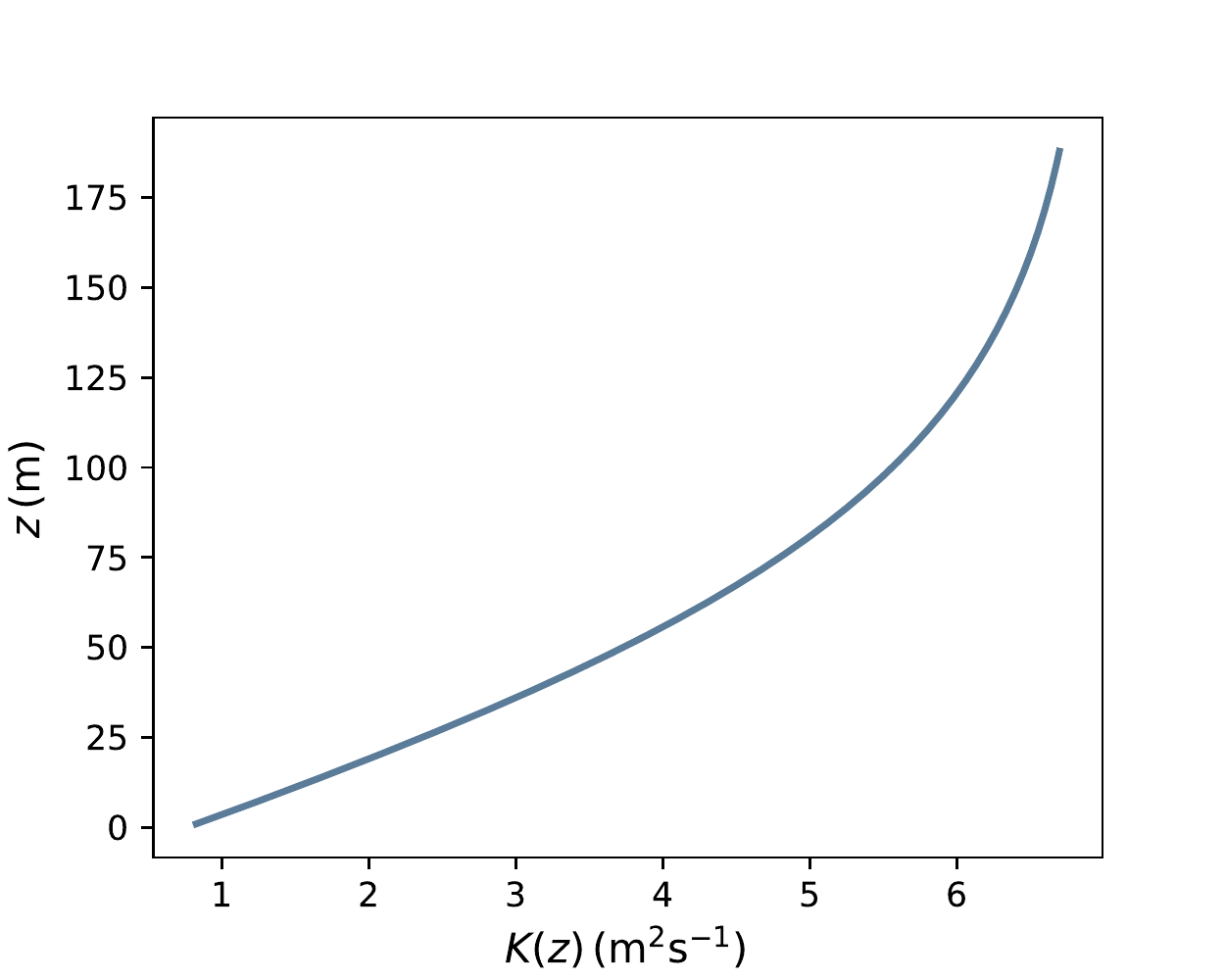}
\includegraphics[width=0.999\textwidth]{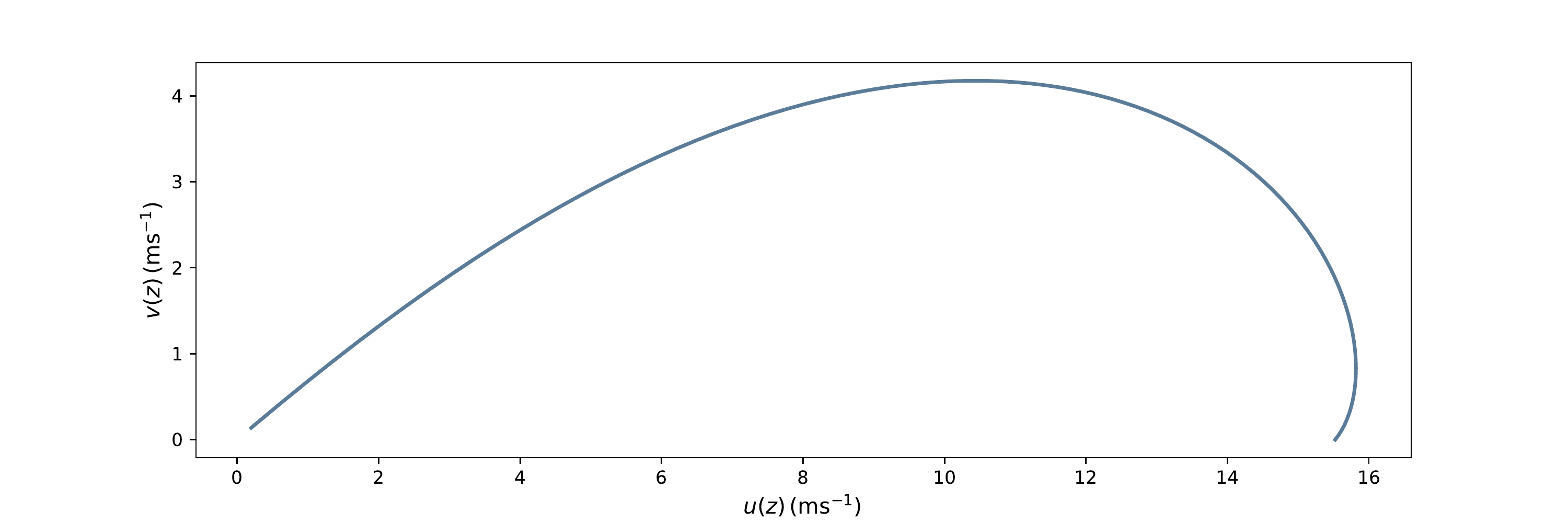}
\caption{ The wind speed $\rho(z)$, relative deflection $\beta(z)$ and eddy viscosity $K(z)$ as well as the hodograph of the \textbf{slow exponential decay model}. These profiles correspond to $\rho_{g}=15\,\mathrm{ms^{-1}}$, $\w=0.02\,\mathrm{ms^{-1}}$ and a maximum relative deflection $\beta_{0}=35^{\circ}$ and at a latitude of $52^{\circ}$ north. Under this choice of parameters, the eddy viscosity coefficient at the bottom of the flow is $K_{0}=0.613\,\mathrm{m^2s^{-1}}$ and the Ekman layer has height $\he\simeq734\,\mathrm{m}$.}\label{fig:slexp}
\end{figure}

Again we define the height of the atmospheric boundary layer as the smallest value $s=\se$ where the condition $\tau(\se)=\tau_{g}$ is first satisfied, with $\se$ obviously being the height of the PBL in terms of the $s-$parameterisation. This condition defines $\se$ implicitly according to
\begin{equation}\label{eg3:se}
0=\pi+\ln\left\vert\frac{\left(\sqrt{1-x}-1\right)\left(\sqrt{1-x}+\sqrt{2}\right)^{\sqrt{2}}}{\left(\sqrt{1-x}+1\right)\left(\sqrt{1-x}-\sqrt{2}\right)^{\sqrt{2}}}\right\vert_{x=e^{-b}}^{x=e^{-(a\se+b)}},
\end{equation}
and since it is not possible to find an exact expression for $\se$ in terms for $a$ and $b$, we revert to numerical methods to determine a value for this height after appropriate values of $a$ and $b$ are chosen. The condition $\w^2>0$  requires $2-\omega^2>0$ and $\omega^4+5\omega^2-2>0$ which restricts the relative deflection angle between the geostrophic wind and the wind at the bottom of the Ekman layer according to $31^\circ <\beta(0) <55^\circ$. The vertical profile of the wind speed, relative deflection, eddy viscosity and hodograph of the wind velocity are shown in figure \ref{fig:slexp}.

\subsection{Square root decay}\label{sec:square_root}
The final case we consider is governed by the separable ODE
\begin{equation}\label{eg4:ode}
 \mu^{\pprime}+\mu^{\prime2}=\left(1+\frac{1}{\sqrt{as+b}}\right)\mu^{\prime2},\quad \mu_{0}^{\prime}=-\frac{a}{2\sqrt{b}}
\end{equation}
where $a$ and $b$ are positive constants. Integrating we find
\begin{equation}\label{eg4:Ms}
\rho(s)=\rho_{g}e^{-\sqrt{as+b}+\sqrt{b}}.
\end{equation}
Equations \eqref{eq:tauprime_separableode} and \eqref{eg4:Ms} yield
\begin{equation}\label{eg4:taus}
 \tau(s)=\tau_{g}+\pi+\left[\frac{1}{2}\ln\left(\frac{\sqrt{x+1}-\sqrt{x}}{\sqrt{x+1}+\sqrt{x}}\right)-\sqrt{x^2+x}\right]_{x=\sqrt{b},}^{x=\sqrt{as+b}}
\end{equation}
where we use the substitution $u=\sqrt{1+\frac{1}{\sqrt{as+b}}}$ to integrate. Meanwhile, equations \eqref{eq:alphaM} and \eqref{eg4:Ms} yield the eddy viscosity coefficient
\begin{equation}\label{eg4:alpha}
 \alpha(s)=\frac{a^2\left(4x^2+6x+1\right)}{8x^3\sqrt{x^2+x}},\quad x=\sqrt{as+b},
\end{equation}
which is clearly positive for all $s>0$ with any choice of positive constants $a$ and $b$. Using the condition $dz=\frac{\alpha(s)}{f}ds$, we may integrate this expression to obtain
\begin{equation}\label{eg4:zs}
 z(s)=\frac{a}{4f}\left[4\ln\left(\left|2\left(\sqrt{x\left(x+1\right)}+x\right)+1\right|\right)-\dfrac{2\sqrt{x+1}\left(16x+1\right)}{3x^\frac{3}{2}}\right]_{x=\sqrt{b}}^{x=\sqrt{as+b}}.
\end{equation}
Given $\frac{dz}{ds}=\frac{\alpha(s)}{f}$ and $\alpha(s)>0$ for all $s>0$, it is clear that $z$ is a monotonically increasing function of $s$, in which case the implicit function theorem ensures there exists a function $s(z)$ which is the inverse of $z(s)$ given above. While in this case no analytic expression is available, we may always calculate its inverse numerically.

\subsubsection{Physical parameters of the square-root decay model}
\begin{figure}[ht!]
\centering
\includegraphics[width=0.333\textwidth]{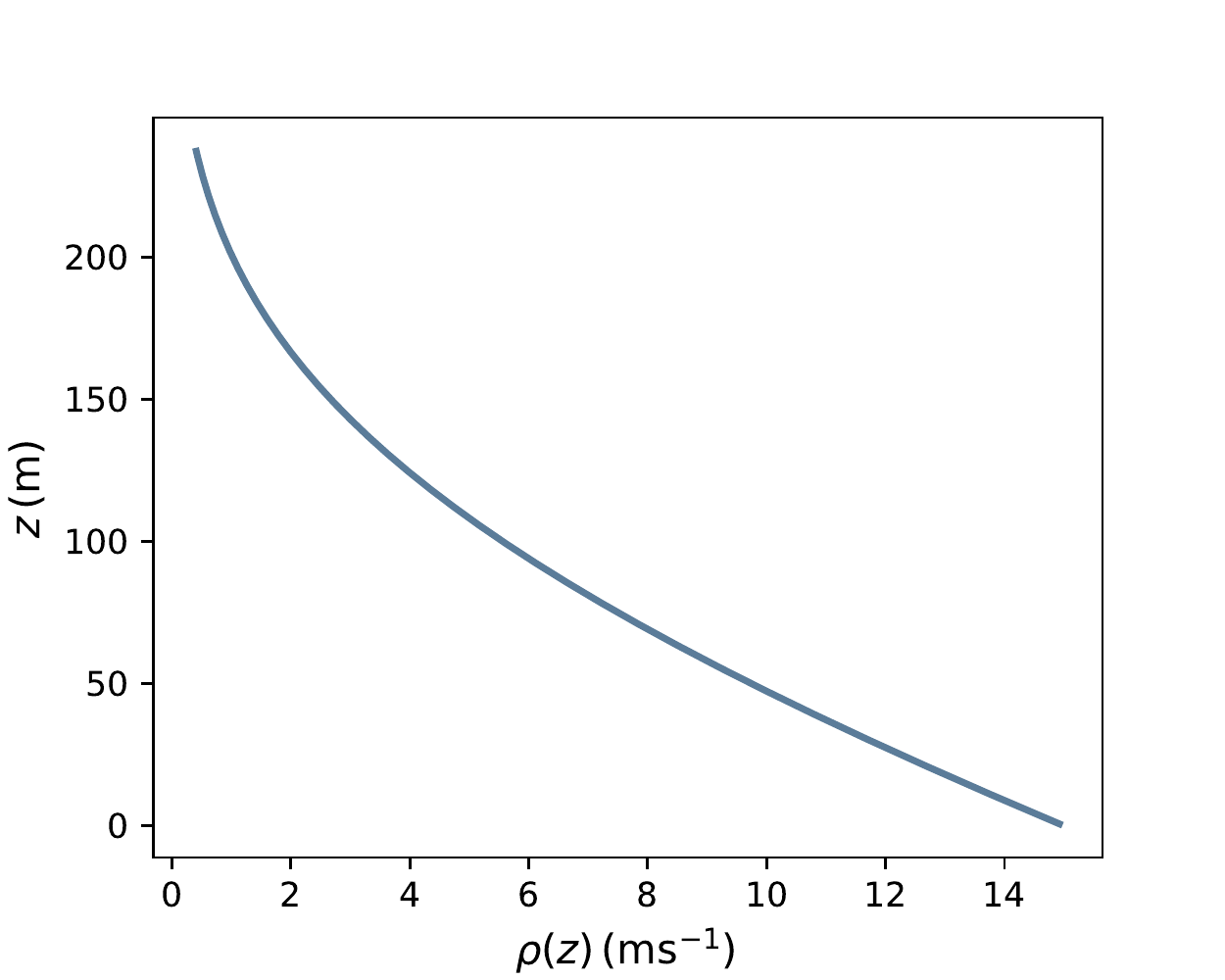}\includegraphics[width=0.333\textwidth]{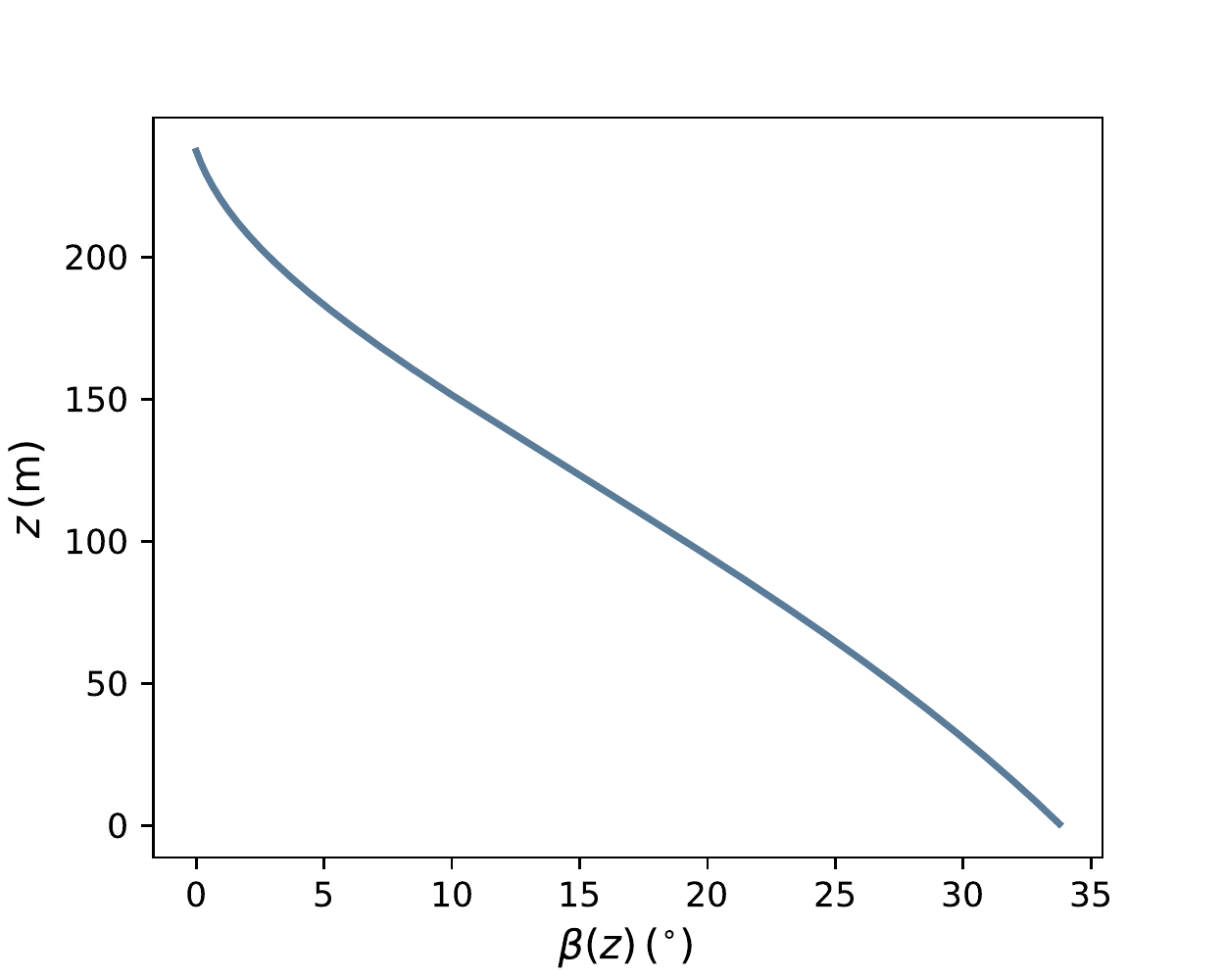}\includegraphics[width=0.333\textwidth,keepaspectratio]{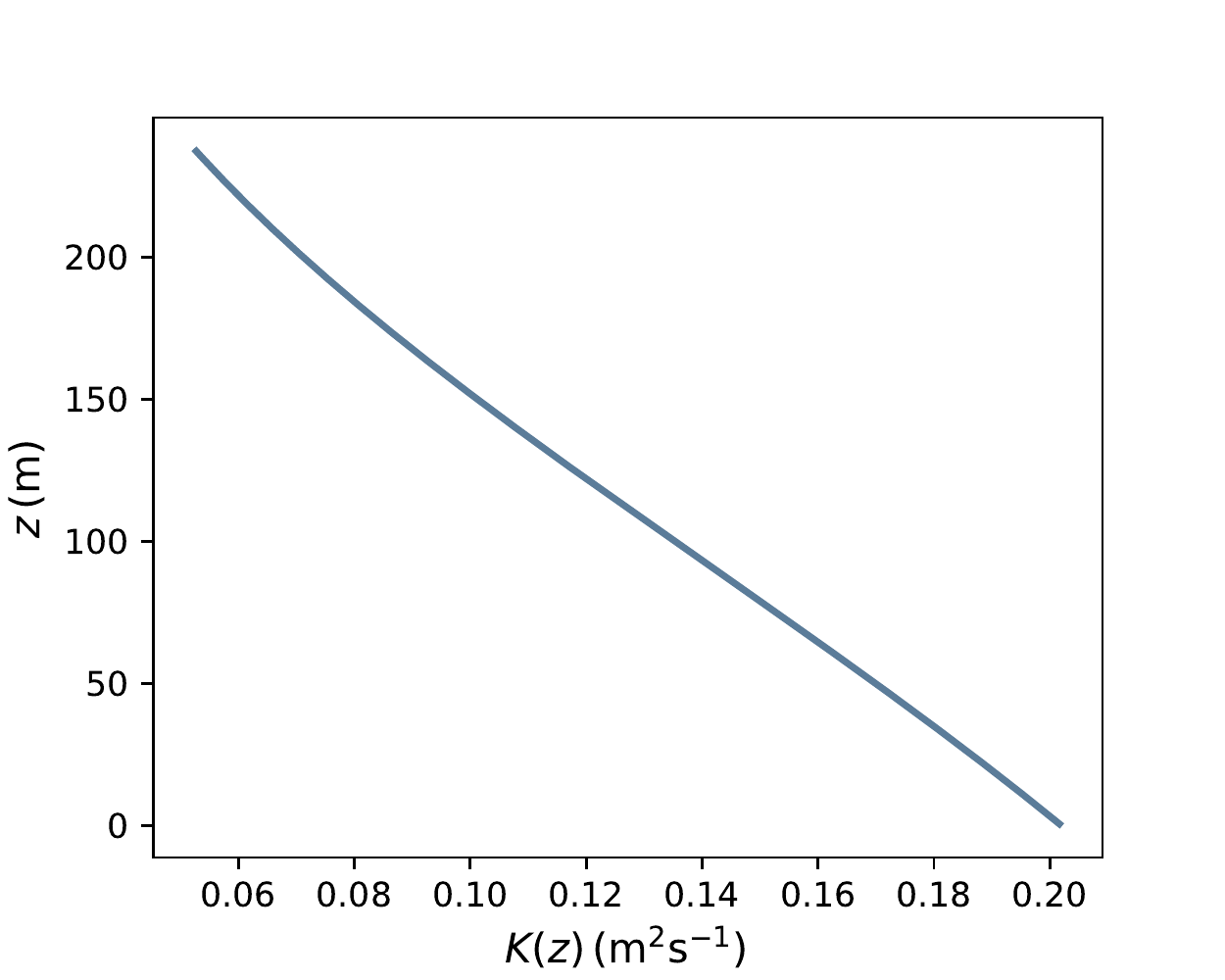}
\includegraphics[width=0.999\textwidth]{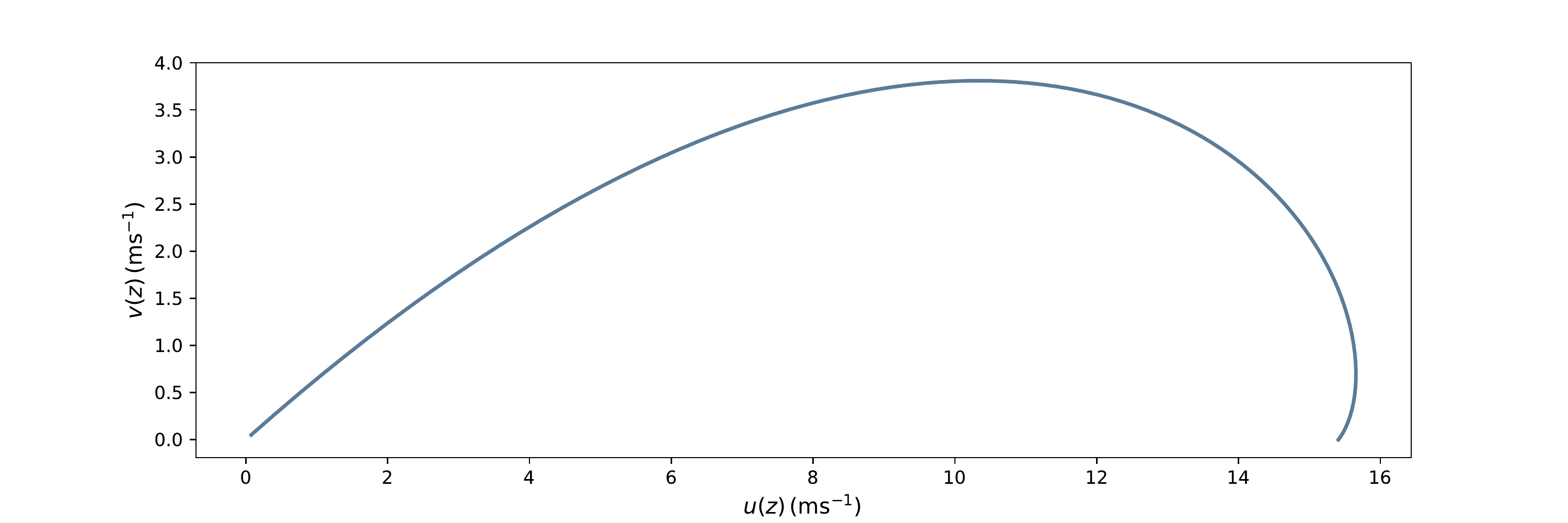}
\caption{ The wind speed $\rho(z)$, relative deflection $\beta(z)$ and eddy viscosity $K(z)$ as well as the hodograph of the \textbf{square-root decay model}. These profiles are obtained from the choice of parameters $\rho_{g}=15\,\mathrm{ms^{-1}}$, $\w=0.02\,\mathrm{ms^{-1}}$ and a maximum relative deflection $\beta_{0}=35^{\circ}$ and at a latitude of $52^{\circ}$ north. Under this choice of parameters, the eddy viscosity coefficient at the bottom of the Ekman layer is $K_{0}=0.614\,\mathrm{m^2s^{-1}}$ and the Ekman layer has height $\he\simeq193\,\mathrm{m}$.}\label{fig:sqrt}
\end{figure}

Using the notation of section \ref{sec:slow_physical} and equations \eqref{eg4:Ms}--\eqref{eg4:taus}, the vertical gradients of $\rho$ and $\tau$ evaluated at $s=0$ satisfy
\begin{equation}
-\frac{a}{2\sqrt{b}}=\frac{K_{0}\dot{\rho}_{0}}{\rho_{g}},\quad-\sqrt{1+\frac{1}{\sqrt{b}}}\frac{a}{2\sqrt{b}}=K_{0}\dot{\rho}_{0},
\end{equation}
which combine to yield
\begin{equation}
 a^2=\frac{4\K\dr\dt}{\rg\omega^2}\equiv\w^2~ \text{and} ~\frac{1}{\sqrt{b}}=\omega^2-1,
\end{equation}
where $\omega=\frac{\rg\dt}{\dr}$ as usual. Using the condition $\alpha(0)=f\K$ and the above relations for $a$ and $b$, we find
\begin{equation}\label{eg4:w}
 \w^2=\frac{8K_{0}f\omega}{\left(\omega^2-1\right)^2\left(\omega^4+4\omega^2-1\right)},
\end{equation}
and so the conditions $\omega^4+4\omega^2-1>0$ and $\omega^2-1<0$ restrict the relative deflection angle between the geostrophic wind  and the ageostrophic wind at the bottom of the PBL according to $26^\circ<\beta(0)<45^\circ$, approximately. This deflection angle, along with the wind speed, the eddy diffusion coefficient and the wind velocity hodograph are shown in figure \ref{fig:sqrt}.

\section{A Riccati equation for the wind speed}
The profiles $\mu(s)$ generated in Section \ref{sec:sepODE} are restricted by the requirement they satisfy a separable ODE of the form \eqref{eq:ODE_separable}. Of course, the requirements for an exponential speed profile of the form $\rho(s)=e^{\mu(s)}$ are $\mu^{\prime}<0$ and $\mu^{\pprime}+\mu^{\prime2}>0$, in which case we may also consider models of the form
\begin{equation}\label{eq:Riccati}
 \mu^{\pprime}(s)+\mu^{\prime}(s)^{2}=\gamma(s),\quad \gamma(s)>0 \text{ for }s>0,
\end{equation}
which is a Riccati type ODE for $\mu^{\prime}(s)$. The reader is also referred to the recent work \cite{Mar2020}, were Riccati equations for the velocity profile $\Phi(s)=e^{\int_{0}^{s}h(\xi)d\xi}$ are explored.  While solutions $\mu^{\prime}(s)$ of equation \eqref{eq:Riccati} are highly contingent on the form of $\gamma(s)$, explicit solutions for $\mu^{\prime}(s)$ are known to exist for an extensive range of functions $\gamma(s)$, see \cite{PZ2017} for instance. However, in contrast to the method presented above, finding an explicit expression for the map $z(s)$ appears to be more challenging under the approach adopted here.

As a simple example, we consider the Riccati equation
 \begin{equation}\label{eq:eg1Riccati}
  \mu^{\pprime}+\mu^{\prime2}=as+b\quad a,b>0
 \end{equation}
where $as+b>0$ for all $s>0$ as required.
Upon applying the Riccati transformation $\mu^{\prime}=\frac{y^{\prime}}{y}$, and introducing the change of variable
$\xi=\frac{1}{\sqrt[3]{a^2}}(as+b)$, equation \eqref{eq:eg1Riccati} now becomes
\begin{equation}\label{eq:eg2_Airy}
 y_{\xi\xi}-\xi y=0,
\end{equation}
where $y_{\xi}$ denotes $\frac{dy}{d\xi}$. Equation \eqref{eq:eg2_Airy} is the well known Airy equation (see \cite{VS2004} for example), whose general solution is of the form
\begin{equation}
 y(\xi)=c_{1}\Ai(\xi)+c_2\Bi(\xi),
\end{equation}
cf. \cite{PZ2017}, where $c_{1,2}$ are arbitrary integration constant  while  $\Ai$ and $\Bi$ are the Airy functions of the first and second kind respectively (see \cite{AS1970}). Given that $\Bi(\xi)$ grows without limit as $\xi\to\infty$, it follows that the physically stable solutions are of the form
\begin{equation}\label{ricc:rho}
\mu(s)=\mu_{g}+\ln\left(\frac{\Ai(\xi(s))}{\Ai(\xi(0))}\right)\Rightarrow \rho(s)=\frac{\rho_{g}\Ai(\xi(s))}{\Ai(\xi(0))}.
\end{equation}
The associated deflection angle is given by
\begin{equation}\label{ricc:tau}
 \tau^{\prime}(s)=-\sqrt{\xi(s)}\Rightarrow \tau(s)=\tau_{g}+\pi-\frac{2\sqrt[3]{a}}{3}\left(\sqrt{\xi(s)^3}-\sqrt{\xi(0)^{3}}\right).
\end{equation}
Given that $\frac{d}{ds}=\frac{1}{\sqrt[3]{a}}\frac{d}{d\xi}$ and $\Ai_{\xi\xi}(\xi)=\xi\Ai(\xi)$, it follows from equation \eqref{eq:alphaM} that
\begin{equation}
 \alpha(s)=-\frac{1}{\sqrt[3]{a}}\left[\frac{1}{2\sqrt{\xi(s)}}+2\sqrt{\xi(s)}\frac{\Ai_{\xi}(\xi(s))}{\Ai(\xi(s))}\right],
\end{equation}
which cannot be integrated in closed form to give $z(s)$. Nevertheless, we may easily perform this integration numerically to determine $z(s)$, and given $\alpha(s)>0$ for all $s>0$, the implicit function theorem (cf. \cite{Die1960}) ensures there exists a unique inverse $s(z)$, which may also be calculated easily using numerical methods.

\subsection*{Acknowledgments}
The author is grateful to the organisers of the workshop ``Mathematical Aspects of Geophysical Flows,'' held at the Erwin Sch\"{o}dinger Institute for Mathematics and Physics, Vienna, Austria, January 20--24, 2020. The author would also like to thank the anonymous referees for several helpful comments.

\subsection*{Conflict of interests}
The author declares there is no conflict of interest with this manuscript.

\end{document}